# Recursive and non-recursive filters for sequential smoothing and prediction with instantaneous phase and frequency estimation applications (extended version)

Hugh Lachlan Kennedy

Hugh.Kennedy@UniSA.edu.au

University of South Australia, STEM Unit, Mawson Lakes Boulevard, Mawson Lakes, SA 5095, Australia.

Hugh.Kennedy@DEWC.com

DEWC Services, Innovation House, Mawson Lakes Boulevard, Mawson Lakes, SA 5095, Australia.

## Abstract

A simple procedure for the design of recursive digital filters with an infinite impulse response (IIR) and non-recursive digital filters with a finite impulse response (FIR) is described. The fixed-lag smoothing filters are designed to track an approximately polynomial signal of specified degree without bias at steady state, while minimizing the gain of high-frequency (coloured) noise with a specified power spectral density. For the IIR variant, the procedure determines the optimal lag (i.e. the passband group delay) yielding a recursive low-complexity smoother of low order, with a specified bandwidth, and excellent passband phase linearity. The filters are applied to the problem of instantaneous frequency estimation, e.g. for Doppler-shift measurement, for a complex exponential with polynomial phase progression in additive white noise. For this classical problem, simulations show that the incorporation of a prediction filter (with a one-sample lead) reduces the incidence of (phase or frequency) angle unwrapping errors, particularly for signals with high rates of angle change, which are known to limit the performance of standard FIR estimators at low SNR. This improvement allows the instantaneous phase of low-frequency signals to be estimated, e.g. for time-delay measurement, and/or the instantaneous frequency of frequency-modulated signals, down to a lower SNR. In the absence of unwrapping errors, the error variance of the IIR estimators (with the optimal phase lag) reaches the FIR lower bound, at a significantly lower computational cost. Guidelines for configuring and tuning both FIR and IIR filters are provided.

# Contents

# 1    Introduction

The measurement of instantaneous phase and/or its first derivative (i.e. frequency) are fundamental operations in sensing systems that rely on wave propagation such as interferometers, spectroscopes, remote-sensors, sounders, rangers, trackers and seekers, over carrier frequencies ranging from Hz to THz, in the fields of aerospace, biomedicine, oceanography, chemistry and astronomy. Without reliable real-time estimates of phase shift (in space for direction or time for radial distance) and frequency shift (for radial velocity), systems such as radars, sonars, and ultrasounds, would not exist and human perception would be limited by the capabilities of our natural senses.

For clarity and simplicity, the treatment here is restricted to the temporal dimension of a (single) sensor or transducer. The instantaneous phase $\theta[n]$ (in radians) of a complex discrete-time sinusoid of unity magnitude is defined using $e^{i\theta[n]}$, where $i$ is the imaginary unit and $n$ is an (integer) sample index into a stream of uniformly sampled sensor measurements, at a rate of $F_s$ (samples per second). For a signal of infinitesimal bandwidth: $\theta[n] = \theta_1 n + \theta_0$, where $\theta_1$ is the frequency (or $\omega$, in radians per sample) and $\theta_0$ is the phase offset (or $\phi$, in radians). For a frequency-modulated signal with non-negligible bandwidth, pulse-compression radars, and echo-locating mammals, transmit and receive waveforms with a swept frequency (e.g. linear or quadratic) for an instantaneous phase that is locally polynomial (e.g. quadratic or cubic) in the sample index. This paper is concerned with the design of online estimators of instantaneous phase derivatives for frequency-swept signals.

Frequency (and phase) estimation methods, for processing the uniformly sampled waveforms produced by sensors and transducers, have a long history and the invention of the Fast Fourier Transform (FFT) was a significant milestone in their development. The FFT is ideal for analysing waveforms comprised of many sinusoidal components *of constant frequency*. However, for a single dominant sinusoid it involves unnecessary computation, especially when high-fidelity estimates are required, because its frequency resolution is inversely proportional to the number of samples processed ($N$) and its computational complexity is proportional to $N \log_2 N$. Furthermore, the latency (and memory requirements) for such block-based (i.e. batch) estimators is much greater than the sample-based (i.e. streaming) estimators considered here.

A straightforward approach to the problem of instantaneous phase estimation for a sampled waveform, such as those produced by a sequence of analogue and digital heterodyning operations or by a Hilbert transformer, is to simply apply the 'argument' operator (also known as the 'angle' operator) to each complex sample, yielding the raw instantaneous phase, followed by a low-pass smoothing filter to attenuate random noise. In principle, instantaneous frequency (e.g. Doppler shift) estimates are then found by numerically computing the first derivative of the instantaneous phase with respect to time, using two-point (backward) differences. However, for non-negligible frequencies and unsophisticated filters, the phase angle wraps around too quickly, and the smoother is unable to track the phase through the angle boundary at $\pm\pi$. (Note that 'angle' is a generic term used here for any radian quantity in the processing chain, such as the phase angle or its rates of change, e.g. its first derivative, i.e. the angular frequency.)

For a sinusoidal signal of constant frequency, taking the derivative *before* the smoother prevents these signal-induced angle-wrap events [1], [2] & [3]. However, differentiators amplify high-frequency noise, thus noise-induced angle-wrap events are now more likely. Moreover, the ability to estimate the phase-offset (e.g. for time-delay or range/distance measurement) is lost. When only frequency estimates are required, the problem of noise amplification is ameliorated by designing the smoother to whiten the coloured noise introduced by the differentiator. This general approach may be extended to handle frequency sweeps (e.g. linear or quadratic) that use more than one differentiator (e.g. two or three [3]) prior to smoothing; however, noise amplification is greater, and low-order phase-derivative information is stripped from the processing chain. It is therefore desirable to minimize the number of pre-smoother differentiation operations and to apply them after the smoother [4]. When this approach is adopted, the smoother must be able to track high rates of angle change without bias and handle frequent angle-wrap events without failure.

Under ideal conditions, the performance of smoothers for phase differences are limited by angle unwrapping errors when the signal-to-noise ratio (SNR) is low and ways of lowering the threshold at which frequency unwrapping fails are explored in [5], [6], [7], [8], [9], [10] & [11]. This is usually accompanied by an increase in computational complexity and/or a decrease in the frequency range of the estimator; however, the simulated scenarios considered here, indicate that optimal (recursive and non-recursive) digital filters reduce the incidence of phase or frequency unwrapping errors, without increasing the computational complexity, or decreasing the effective bandwidth, of the estimator.

Estimator performance at high SNR may be limited by the presence of multiple swept-frequency components of variable duration/scale with phase polynomials of unknown degree. Consideration of these cases with multiple unknown parameters in iterative (e.g. maximum likelihood) algorithms, further increases computational complexity [12], [13], [14], [15] & [16], which limits their application to the post-processing of captured data records. Off-line batch methods are not constrained by causality, as data from the distant past and distant future are available at any given sample, whereas the number of 'future' samples available in on-line streaming methods is limited to the passband group-delay parameter $q$. Both approaches typically rely on polynomial regression operations to estimate phase/frequency parameters from an unwrapped sequence of angle measurements.

Non-recursive digital filters with a Finite Impulse Response (FIR) are usually used for the smoothing function in online estimators of instantaneous frequency. They may be regarded as a simple sliding-batch algorithm. At high SNR, the error variance of such filters reaches the Cramer-Rao lower-bound (CRLB); however, long impulse responses are required to reduce the error variance of the frequency estimator. The computational complexity of FIR filters is proportional to the impulse response duration ($M$, in samples). Polyphase filter-banks utilize sample-rate reduction to lower the computational complexity of long FIR filters with step-like low-pass responses [17], [18] & [19]. However, the optimal filters for this problem have relatively smooth (non-oscillatory) impulse responses that are not strictly bandlimited; furthermore, lowered sampling frequencies increases the per-sample rate of instantaneous phase change, thus the likelihood of unwrapping errors.

The computational complexity of recursive digital filters with an Infinite Impulse Response (IIR) is independent of the impulse response duration. However, conventional procedures for their design are not straightforward because they traditionally involve the determination of both the zeros and the poles of the filter's discrete-time transfer function (for FIR filters, all poles are fixed at the origin of the complex z-plane). Constraints must also be applied to ensure that all poles are well within the unit circle (for robust stability). Thus, elaborate optimizers, that are customized to suit design priorities are usually used for the determination of the filter coefficients [20], [21], [22] & [26]. However, features of the filter response that are of high priority for generic operations in digital signal processing (DSP) components such as a flat passband with nearly linear phase, a flat stopband with negligible gain, with a transition of negligible width in between [20], [22] & [23], are not critical for the problem of instantaneous phase and/or frequency estimation as it is posed here.

Using simple two-point differences to compute instantaneous phase derivatives may seem like a naïve oversimplification; however, similar to the approach adopted in [23] & [24], individual components (e.g. the optimal low-pass filter) are designed by considering the integrated system with all filters (e.g. the simple differentiators) in the processing path. The recent literature on IIR differentiator design deals mainly with wideband differentiators [21], [25] & [26]; however, these filters have compact impulse responses which may be efficiently synthesized using simpler FIR filter designs; moreover, they provide little or negligible high-frequency attenuation, which is not ideal for processing noisy sensor signals, without an additional low-pass filtering stage.

Recursive Bayesian approaches, such as the Kalman filter, are a possible design alternative [27]; however, in the scenarios considered here, there is no process noise, the variance of the measurement noise is unknown, and it is not necessarily Gaussian, thus there is limited prior knowledge to be leveraged. The coefficients of the alpha-beta smoothing filter considered in [2] may be derived using the steady-state gain of a second order (i.e. constant angular velocity) Kalman filter; however, these second-order smoothers have insufficient degrees of freedom to accommodate all of the design parameters considered here.

The procedure for the design of the low-pass IIR filters (for estimation and prediction) proposed in this paper, minimises the noise colour introduced by the differentiators in the processing chain, subject to derivative constraints at the origin in the frequency domain, thus moment constraints around the phase origin in the time domain, to ensure that an angular polynomial of specified degree is tracked without bias at steady state. The solution is expressed as a linear combination of first-order basis-functions. In the general IIR case, the basis-functions are damped oscillators. The poles of the basis-set correspond to the complex poles of a discretised analogue prototype with the desired bandwidth. A Bessel filter prototype is recommended for good passband phase linearity. The use of predetermined filter poles to broadly satisfy bandwidth and stability requirements greatly simplifies the design procedure. The zeros of the discrete-time filter transfer-function are determined from the linear coefficients, which hone the solution to satisfy the specific requirements of the estimation problem. This linear approach, with specified poles, greatly simplifies the problem of feedback filter design, which is usually posed as a non-linear optimization, with unknown poles. It has previously been used by the author to design target trackers & pulse detectors (for real signals) [27], feedback controls [28], and non-causal filters for feature detection/classification in images [29].

For both FIR and IIR low-pass filters, a long passband group delay is required to minimize the noise gain, thus the error variance of an unbiased *estimator* (Note that these low-pass estimators with a phase lag are also referred to as 'smoothers' in this paper). Linear-phase FIR filters, with an impulse response duration of $M$ samples, have a group delay of $q = (M-1)/2$; for IIR filters, selection of the optimal group delay ($q$) is not so straightforward; however, a novel procedure for its determination is incorporated into the design procedure outlined above and detailed herein. Furthermore, when low-pass filtering high rates of angular change, use of the lagged estimate degrades angle unwrapping decisions, thus a low-pass *predictor* with a one sample lead ($q = -1$) for unwrapping purposes, that works in tandem with the estimator, is also recommended here.

The main technical contributions of this paper are summarized in the next section. An overview of the assumptions and theory is provided in Section 3, then the proposed procedure for the design of suitable FIR and IIR filters for phase and/or frequency estimation is described in Section 4. Linear state-space recursions for the realization of recursive IIR filters, incorporating the estimator (for noise attenuation) and the predictor (for phase unwrapping) are described in Section 5. The performance of various filters is then analysed and discussed using Monte Carlo simulations in Section 6 and Section 7, followed by some closing remarks in Section 8.

## Acronyms

| | |
|---|---|
| ADC | Analogue-to-Digital Converter |
| BPF | Band-Pass Filter |
| CNG | Coloured-Noise Gain |
| CRLB | Cramer-Rao Lower-Bound |
| dc | direct current (0 Hz) |
| DSP | Digital Signal Processing |
| FFT | Fast Fourier Transform |
| FIR | Finite Impulse Response |
| HPF | High-Pass Filter |
| IIR | Infinite Impulse Response |
| LPF | Low-Pass Filter |
| LSS | Linear State-Space |
| MC | Monte Carlo (simulation) |
| PSD | Power Spectral Density |
| SNR | Signal to Noise Ratio |
| WNG | White-Noise Gain |

## Symbols

| | |
|---|---|
| $i$ | Imaginary unit |
| $s$ | Coordinate in the complex s-plane, reached via the $\mathcal{L}$ (i.e. Laplace) transform |
| $z$ | Coordinate in the complex z-plane, reached via the $\mathcal{Z}$ transform |
| $\omega$ | Relative angular frequency (radians per sample) |
| $f$ | Relative normalized frequency (cycles per sample) |
| $\Omega$ | Angular frequency (radians per second) |
| $F$ | Frequency (cycles per second or Hz) |
| $F_s$ | Sampling frequency (samples per second or Hz) |
| $T_s$ | Sampling period (seconds per sample) |
| $n$ | Sample index, $n = 0 \dots N-1$ |
| $\psi[n]$ | Complex sampled signal |
| $A$ | Magnitude of the signal, which is assumed to be a (real) constant |
| $\varepsilon_\psi[n]$ | White noise (complex) added to the (complex) signal |
| $\sigma_\varepsilon^2$ | Variance of $\varepsilon_\psi$ |
| $x[n]$ | Raw digitized (complex) waveform to be processed, signal plus noise |
| $\theta[n]$ | Instantaneous phase (real), phase angle, or simply the phase (radians) of the signal |
| $\tilde{\theta}[n]$ | Raw (i.e. noise corrupted) phase-angle measurements |
| $\bar{\theta}[n]$ | Unwrapped phase-angle measurements |
| $\hat{\theta}[n]$ | Instantaneous phase estimate |
| $\varepsilon_\theta[n]$ | Phase noise (real), transformed from the complex domain |
| $K_\theta$ | Order or degree of the signal's instantaneous phase polynomial |
| $k_\theta$ | Monomial index of the instantaneous phase polynomial, for $k_\theta = 0 \dots K_\theta$ |
| $\theta_{k_\theta}$ | The $k_\theta$th coefficient of the instantaneous phase polynomial |
| $\theta_0$ | Phase offset of the signal (radians) |
| $\theta_1$ | Phase velocity of the signal (radians per sample) |
| $\theta_2$ | Phase acceleration of the signal (radians per sample per sample) |
| $\theta_3$ | Phase jerk of the signal (radians per sample per sample per sample) |
| $\dot{\theta}$ | First derivative of the instantaneous phase, with respect to time |
| $\ddot{\theta}$ | Second derivative of the instantaneous phase, with respect to time |
| $\dddot{\theta}$ | Third derivative of the instantaneous phase, with respect to time |
| $\phi$ | Phase offset of the signal (radians), $\phi = \theta_0$ |
| $\omega_\psi$ | Instantaneous frequency of the signal (radians per sample), $\omega_\psi = \dot{\theta}$ |
| $f_\psi$ | Instantaneous frequency of the signal (cycles per sample) |
| $F_\psi$ | Instantaneous frequency of the signal (cycles per second or Hz) |
| $\Delta\theta[n]$ | Backward two-point phase-difference (radians) at the time of the $n$th sample |
| $L$ | Number of lags |
| $l$ | Lag index, $l = 0 \dots L-1$ |
| $\tilde{\omega}[n]$ | Raw phase-angle-differences or raw frequency measurements |
| $\bar{\omega}[n]$ | Unwrapped frequency measurements |
| $\hat{\omega}_\psi[n]$ | Instantaneous frequency estimate |
| $\varepsilon_\omega[n]$ | Frequency noise |
| $y[n]$ | Output of the filtering operation at the time of the $n$th sample |
| $M$ | Window length (samples) |
| $m$ | Delay index (samples), $m = 0 \dots M-1$ |
| $h_{\mathrm{LPF}}[m]$ | Impulse response of the low-pass filter |
| $H_{\mathrm{LPF}}(e^{i\omega})$ | Frequency response of the low-pass filter |
| $q$ | Passband group delay (samples) of the low-pass filter |
| $P_{\varepsilon_\omega}(\omega)$ | Power spectral density of the coloured noise |
| $h_{\mathrm{dif}}[m]$ | Impulse response of a two-point differentiator |
| $H_{\mathrm{dif}}(z)$ | Transfer function of a two-point differentiator |
| $H_{\mathrm{dif}}(e^{i\omega})$ | Frequency response of a two-point differentiator |
| $h_{\mathrm{HPF}}[m]$ | Impulse response of the high-pass filter |
| $H_{\mathrm{HPF}}(e^{i\omega})$ | Frequency response of the high-pass filter |
| $K_1$ | Number of monomial regressors for a fitted polynomial with a degree of $K_1 - 1$ |
| $k_1$ | Order of dc derivative, or monomial regressor index, $k_1 = 0 \dots K_1 - 1$ |
| $K_0$ | Number of two-point differentiators in the processing chain, as assumed during filter design |
| $\boldsymbol{h}$ | $M \times 1$ vector with elements $h_{\mathrm{LPF}}[m]$ |
| $\boldsymbol{X}$ | $M \times K_1$ (Vandermonde) matrix with the regressor monomials as its columns |
| $\boldsymbol{x}(q)$ | $1 \times K_1$ 'synthesis' vector, that evaluates the locally fitted polynomial at $n - q$ |
| $\boldsymbol{W}$ | $M \times M$ 'whitening' matrix |
| $\boldsymbol{P}_{\varepsilon_\omega}$ | $M \times M$ (Toeplitz) covariance matrix of the coloured noise |
| $r_{\varepsilon_\omega}[l]$ | Discrete-time coloured-noise autocorrelation function |
| $\sigma_\omega^2$ | Variance of the frequency estimation error |
| $v_{\text{LPF}}$ | White-noise gain of the LPF |
| $v_{\text{BPF}}$ | White-noise gain of the BPF, i.e. the coloured-noise gain of the LPF |
| $h_{\text{BPF}}[m]$ | Impulse response of the band-pass filter |
| $H_{\text{BPF}}(e^{i\omega})$ | Frequency response of the band-pass filter |
| $\boldsymbol{I}$ | Identity matrix |
| $\sigma_\theta^2$ | Variance of the phase estimation error |
| $K_\varphi$ | Number of first-order basis-functions used to define the response of the LPF |
| $k_\varphi$ | Basis-function or internal-state index (for $k_\varphi = 0 \dots K_\varphi - 1$) or IIR filter coefficient index (for $k_\varphi = 0 \dots K_\varphi$) |
| $c_{k_\varphi}$ | Linear coefficient (complex) corresponding to the $k_\varphi$th basis-function |
| $p_{k_\varphi}$ | Pole of the $k_\varphi$th basis function in the complex z-plane |
| $\tilde{p}_{k_\varphi}$ | $k_\varphi$th continuous-time pole of the analogue prototype filter in the complex s-plane |
| $\varphi_{k_\varphi}[m]$ | Impulse response of the $k_\varphi$th basis function |
| $\varphi_{k_\varphi}(z)$ | Transfer function of the $k_\varphi$th basis function |
| $\varphi_{k_\varphi}(e^{i\omega})$ | Frequency response of the $k_\varphi$th basis function |
| $\boldsymbol{c}$ | $K_\varphi \times 1$ vector with the $c_k$ coefficients as elements |
| $\boldsymbol{S}$ | $K_\varphi \times K_\varphi$ Hermitian matrix of definite integrals containing the noise power spectral density and pairwise basis-function products |
| $S_{k_m k_n}$ | Element in the $k_m$th row and $k_n$ column of $\boldsymbol{S}$ |
| $g_k[n]$ | High-pass filtered basis-function impulse response, used to evaluate $S_{k_m k_n}$ in the time domain |
| $\boldsymbol{\lambda}$ | $K_1 \times 1$ vector containing the Lagrange multipliers |
| $\boldsymbol{\Phi}$ | $K_1 \times K_\varphi$ matrix containing the dc derivatives of the basis functions |
| $\Phi_{k_1,k_\varphi}$ | Element in the $k_1$th row and $k_\varphi$th column of $\boldsymbol{\Phi}$ |
| $\boldsymbol{d}(q)$ | $K_1 \times 1$ vector containing the desired dc derivatives as a function of $q$ |
| $d_{k_1}$ | Element in the $k_1$th row of $\boldsymbol{d}(q)$ |
| $p$ | Repeated real pole in the complex z-plane |
| $v'_{\text{BPF}}(q)$ | Derivative of $v_{\text{BPF}}(q)$ with respect to $q$ |
| $B(z)$ | Numerator polynomial of $H_{\text{LPF}}(z)$ |
| $A(z)$ | Denominator polynomial of $H_{\text{LPF}}(z)$ |
| $X(z)$ | $\mathcal{Z}$-transform of low-pass filter input |
| $Y(z)$ | $\mathcal{Z}$-transform of low-pass filter output |
| $b[k_\varphi]$ | $k_\varphi$th coefficient of the $B(z)$ polynomial |
| $a[k_\varphi]$ | $k_\varphi$th coefficient of the $A(z)$ polynomial |
| $\boldsymbol{w}[n]$ | $K_\varphi \times 1$ state vector of a discrete-time LSS system at the time of the $n$th complex sample |
| $\boldsymbol{G}$ | $K_\varphi \times K_\varphi$ state-transition matrix of a discrete-time LSS system |
| $\boldsymbol{H}$ | $K_\varphi \times 1$ input vector of a discrete-time LSS system |
| $\boldsymbol{C}$ | $1 \times K_\varphi$ output vector of a discrete-time LSS system |
| $\boldsymbol{C}_{\text{est}}$ | Output vector of the LSS estimator |
| $\boldsymbol{C}_{\text{prd}}$ | Output vector of the LSS predictor |
| $\hat{x}[n]$ | Predicted angle input of the next complex sample |
| $\tilde{x}[n]$ | Wrapped angle input of the $n$th complex sample |
| $\bar{x}[n]$ | Unwrapped angle input of the $n$th complex sample |
| $\boldsymbol{w}_0$ | $K_\varphi \times 1$ state initialization vector |
| $N$ | Number of samples in a batch or block for processing |
| $\alpha$ | Number of (pre) differentiators placed before the LPF |
| $\beta$ | Number of (aft) differentiators placed after the LPF |
| $f_0$ | Instantaneous frequency of signal at the start of a simulation batch |
| $f_1$ | Instantaneous frequency of signal at the centre of a simulation batch |
| $f_2$ | Instantaneous frequency of signal at the end of a simulation batch |
| $\widetilde{K}_0$ | Number of two-point differentiators in the processing chain for a given system configuration in an MC simulation |
| $\tilde{v}_{\text{BPF}}$ | Coloured-noise gain of the LPF for a given system configuration |

## 2   Executive summary

Estimates of a single sinusoid's instantaneous phase or frequency, computed by convolving a (uniformly sampled) waveform, with a well-designed digital (low-pass or band-pass) filter, are precise at high SNR and accurate when the following assumptions are met: the (heterodyned or Hilbert transformed) waveform contains only one (complex) sinusoidal signal, the signal magnitude is constant, and when the (complex) additive noise is white. The phase-angle of the frequency-swept sinusoid (a complex exponential) is assumed to progress as a low-order polynomial and (high-pass) differentiators are usually used to constrain the (low-pass) filtered angle to $\pm\pi$. However, phase-unwrapping errors, due to the random noise and/or the phase-modulated signal, are known to cause such simple methods to fail at low SNR (around 10 dB). A procedure that applies a phase-lag for estimation, and a phase-lead for prediction (to be used for phase unwrapping), is recommended here and shown to lower the failure threshold (down to around 3 dB in some cases). Non-recursive digital filters with a finite impulse response (FIR) are usually used but it is shown here that recursive filters with an infinite impulse response (IIR) provide similar or better performance at a lower computational cost. A simple design-procedure that minimizes the coloured-noise gain of the filter, subject to $K_1$ flatness constraints at zero frequency is described. The white noise is coloured by the $K_0$ digital differentiators in the processing chain. The $K_1$ flatness constraints ensure that at high SNR (where phase unwrapping errors are unlikely) the $K_0$th derivative of a $K_\theta$th-degree polynomial is tracked without bias at steady state (when $K_1 > K_\theta - K_0$). The passband group-delay $q$, is solved to minimize the error-variance of the estimator, and of the $2(K_1 - 1) - 1$ feasible solutions, the one with the best passband phase linearity is selected. The delay of the predictor is set using $q = -1$. The proposed design procedure yields the more well-known linear-phase FIR filters as a limiting case. Monte-Carlo simulations are used to compare the suitability of both FIR and IIR solutions, in hypothetical instruments that estimate instantaneous phase (e.g. for time-delay measurements, using $K_0 = 0$) and instantaneous frequency (e.g. for Doppler-shift measurements, using $K_0 = 1$) for a sinusoidal signal with cubic instantaneous phase ($K_\theta \leq 3$). The performance of the FIR and IIR filters is almost identical, and the observed error-variance agrees with the theoretical values that are evaluated using the coloured noise-gain of each filter. The non-recursive FIR filters are realized using 64 delay-multiply-add operations; whereas the (fifth order) recursive IIR filters are realized using only 10, which makes them an attractive architectural alternative in miniature embedded computers operating at MHz or GHz rates.

## 3   Mathematical preliminaries

The sampled waveform is modelled as follows:

$$x[n] = \psi[n] + \varepsilon_\psi[n] \text{ where} \tag{3.1}$$

$x[n]$ is the raw digitized (complex) *waveform* to be processed, a so-called analytic signal, sampled at a uniform rate of $F_s$ (samples per second or Hz) for a sampling period of $T_s = 1/F_s$ (seconds per sample)

$n$ is the sample index

$\varepsilon_\psi[n]$ is (complex) white noise with real and imaginary parts 'drawn' from a random distribution (not necessarily Gaussian) with a mean of zero and a variance of $\sigma_\varepsilon^2$ and

$\psi[n]$ is a complex *signal*.

The signal is further assumed to have the following form:

$$\psi[n] = Ae^{i\theta[n]} \text{ where} \tag{3.2}$$

$i$ is the imaginary unit

$e$ is Euler's number

$A$ is the magnitude of the signal, which is assumed to be a (real) constant

$\theta[n]$ is the instantaneous phase, phase angle, or simply the *phase* (in radians) of the signal.

The angle argument of the exponential is obtained by applying the $\arg\{\blacksquare\}$ operator, also known as the angle operator $\angle$ (they are used interchangeably here) i.e.

$$\angle x[n] = \tilde{\theta}[n] = \theta[n] + \varepsilon_\theta[n] \text{ where} \tag{3.3}$$

$\tilde{\theta}[n]$ are the raw (i.e. noise corrupted) phase-angle measurements and

$\varepsilon_\theta[n]$ is the phase noise with

$\varepsilon_\theta[n] \approx \varepsilon_\psi[n]/A$

at high SNR, i.e. when $\left(|\psi|/|\varepsilon_\psi|\right)^2 \gg 1$ thus $|\varepsilon_\psi/A| \ll 1$ so that $\tan^{-1}\{\varepsilon_\psi/A\} \approx \varepsilon_\psi/A$, where $|\blacksquare|$ is the magnitude operator. At low SNR the noise distribution is not so simple [30] & [31], and unmodelled correlations/distortions increase the likelihood of phase unwrapping errors.

The instantaneous phase is assumed to rotate or progress as a low-order polynomial

$$\theta[n] = \sum_{k_\theta=0}^{K_\theta} \theta_{k_\theta} n^{k_\theta}/k_\theta! \text{ where} \tag{3.4}$$

$K_\theta$ is the order or degree of the instantaneous phase polynomial and

$\theta_{k_\theta}$ is the $k_\theta$th polynomial coefficient.

For non-polynomial signals, local (low order) polynomial approximations (e.g. Taylor series) are usually a reasonable approximation, over sufficiently short timescales.

For a linear model, i.e. a first-order polynomial with $K_\theta = 1$, the phase rotates at a constant rate as

$$\theta[n] = \theta_1 n + \theta_0 \text{ where} \tag{3.5}$$

$\theta_0$ is the phase offset $\phi$ of the signal (radians) and

$\theta_1$ is the phase velocity of the signal (radians per sample).

The instantaneous frequency $\omega_\psi$ (radians per sample) is defined as the first derivative of the instantaneous phase with respect to time (i.e. $\dot{\theta}$), thus for this linear polynomial it is constant, i.e.

$$\omega_\psi = \dot{\theta} = \theta_1. \tag{3.6}$$

For a quadratic model (with $K_\theta = 2$) the $\theta_2$ parameter is the second derivative of phase with respect to time (i.e. $\ddot{\theta} = \theta_2$), which is referred to here as the phase acceleration, or linear frequency sweep rate, or simply the *chirp* (radians per sample per sample), thus

$$\omega_\psi[n] = \dot{\theta} = \theta_2 n + \theta_1. \tag{3.7}$$

For a cubic model (with $K_\theta = 3$) the $\theta_3$ parameter is the third derivative of phase with respect to time (i.e. $\dddot{\theta} = \theta_3$), which is referred to here as the phase jerk (radians per sample per sample per sample), thus

$$\omega_\psi[n] = \dot{\theta} = \theta_3 n^2/2 + \theta_2 n + \theta_1. \tag{3.8}$$

Note that as an alternative to the *relative angular frequency* $\omega$ (radians per sample), the *relative normalized frequency* $f$ (cycles per sample) is also used here, with $f = \omega/2\pi$. These frequencies are relative to the sampling rate, and they are converted to angular frequency (radians per second) and frequency (cycles per second or Hz) using $\Omega = F_s\omega$ and $F = F_s f$ and, respectively. The $\blacksquare_\psi$ subscript is used here to distinguish the frequency parameters (e.g. $\omega_\psi$ or $f_\psi$) of the sinusoidal waveform from the frequency coordinate in the transform domain (e.g. $\omega$ or $f$).

The phase angle is confined to the circular $(-\pi, \pi]$ interval, thus the polynomial 'wraps around' as it passes through the $\pm\pi$ boundary. For a constant-frequency signal, in the absence of noise, periodic phase-wrap events are avoided by applying the angle operator to delayed conjugate products of the input sequence, i.e. $\arg\{x[n]x^*[n-l]\}$, where $l = 1$ and $\blacksquare^*$ denotes complex conjugation.

When processing the delayed conjugate products, in the absence of noise, such that $x[n] = \psi[n]$, we have

$$\arg\{\psi[n]\psi^*[n-1]\} = \arg\{A^2 e^{i\theta[n]} e^{-i\theta[n-1]}\} = \arg\{A^2 e^{i\theta[n]-i\theta[n-1]}\} \qquad (3.9a)$$

$$= \theta[n] - \theta[n-1] = \Delta\theta[n] \text{ where} \qquad (3.9b)$$

$\Delta\theta[n]$ is the (backward) phase difference at the time of the $n$th sample.

For a constant frequency (i.e. a phase polynomial with $K_\theta = 1$) substitution of (3.5) into (3.9b) yields

$$\Delta\theta[n] = \theta[n] - \theta[n-1] = \theta_1 n + \theta_0 - \{\theta_1(n-1) + \theta_0\} = \theta_1 = \omega_\psi . \qquad (3.10)$$

In the absence of noise, operating on the delayed conjugate products $\psi[n]\psi^*[n-1]$, eliminates the need for unwrapping when the frequency of the signal is constant ($K_\theta = 1$) because the output of the angle operator is a constant (the frequency to be estimated) and does not exceed $\pm\pi$. However, unwrapping may be required for the $x[n]x^*[n-1]$ sequence when the variance of the noise ($\sigma_\varepsilon^2$) is large and when the $F_\psi$ approaches $\pm F_s/2$ (where $F_\psi$ is the signal frequency), because random perturbations may cause the output of the angle operator to exceed $\pm\pi$.

Generalizations of this approach that consider a wider range of delays by operating on the discrete-time auto-correlation function, i.e. using $l = 1 \dots L - 1$, have also been proposed. However, they are not considered here because the frequency range of these estimators is drastically reduced from $\pm\pi$ [1] to $\pm\pi/L$ [6] and $\pm 2\pi/(L+1)$ [7], respectively. These methods are claimed to reduce the variance of the frequency estimate at low SNR.

For chirped exponentials with a linear frequency sweep (i.e. a quadratic phase progression with $K_\theta = 2$) dual two-point differences may be cascaded in series, which is realized using $\arg\{x[n]x^*[n-1]x^*[n-2]x[n-3]\}$ to prevent the argument of the exponential from wrapping around, in the absence of noise [3]. These second differences yield the rate of frequency change (plus noise). However, this technique is not considered here because it is assumed that estimates of frequency (shift), and optionally phase (delay), are required.

*In the presence of noise*, (3.1) may be re-written as

$$x[n] = Ae^{i\theta[n]} + \varepsilon_\psi[n] = Ae^{i\theta[n]+i\varepsilon_\theta[n]} \text{ thus} \qquad (3.11a)$$

$$\arg\{x[n]x^*[n-1]\} = \widetilde{\omega}[n] = \omega_\psi + \varepsilon_\omega[n] \text{ where} \qquad (3.11b)$$

$\widetilde{\omega}[n]$ are raw phase-angle-differences or raw frequency measurements and

$\varepsilon_\omega[n]$ is the frequency noise.

The raw frequency measurements $\widetilde{\omega}[n]$, are then sequentially unwrapped, yielding $\overline{\omega}[n]$.

Smoothed estimates of first phase derivative $\dot{\theta}$, are computed by regressing a polynomial model to the unwrapped sequence $\overline{\omega}[n]$. The regression is applied over a sliding window using

$y[n] = \sum_{m=0}^{M-1} h_{\text{LPF}}[m]\bar{\omega}[n-m]$, where $\qquad$ (3.12)

$y[n]$ is the output of the smoothing operation at the *current* time of the $n$th sample

$m$ is the delay index (samples)

$M$ is the window length (samples) and

$h_{\text{LPF}}[m]$ is the impulse response of the smoother, which is a low-pass filter (LPF), with frequency response $H_{\text{LPF}}(e^{i\omega})$.

For an FIR LPF (with $M < \infty$) the convolution is applied non-recursively. For an IIR LPF (with $M = \infty$) the convolution is applied recursively (see Section 5).

For a signal with a constant frequency ($K_\theta = 1$) the convolution is simply a sliding weighted-average operation that yields a 'smoothed' estimate of the $\theta_1$ parameter, i.e. $\hat{\theta}_1 = \hat{\omega}_\psi$. For signals with non-constant or 'swept' frequency (i.e. $K_\theta > 1$) the latency of these smoothing filters should be considered, and their outputs interpreted as follows:

$y[n] = \hat{\omega}_\psi[n-q]$ where $\qquad$ (3.13)

$\hat{\omega}_\psi[n-q]$ is a *delayed* estimate of the signal's frequency at the time of the $(n-q)$th sample and

$q$ is the passband group delay (samples) of the filter. For low-pass (FIR & IIR) digital filters, it is the negative of the derivative of the phase response, i.e. $\angle H_{\text{LPF}}(e^{i\omega})$, with respect to angular frequency, evaluated at the dc limit. For linear-phase FIR filters $q = (M-1)/2$.

Kay points out that when $\varepsilon_\psi[n]$ is white, the noise sequence $\varepsilon_\omega[n]$ *is not white*, after the angle operator is applied to the delayed conjugate products $x[n]x^*[n-1]$. Fortunately, the Power Spectral Density (PSD) of the coloured noise $P_{\varepsilon_\omega}(\omega)$, is readily derived in the high-SNR case. During the transformation from delayed conjugate products (a complex signal) to phase differences $\tilde{\omega}[n]$ (a real polynomial), the white-noise sequence $\varepsilon_\theta[n]$ effectively passes through a first-order system (a two-point numerical differentiator) with impulse response,

$h_{\text{dif}}[m] = \begin{bmatrix} 1 & -1 \end{bmatrix}$ $\qquad$ (3.14a)

which is a digital high-pass filter (HPF) with the following transfer function:

$H_{\text{dif}}(z) = \frac{z-1}{z} = 1 - 1/z$ thus $\qquad$ (3.14b)

$P_{\varepsilon_\omega}(\omega) = \left| H_{\text{dif}}(e^{i\omega}) \right|^2 = \left| 1 - e^{-i\omega} \right|^2$. $\qquad$ (3.14c)

The 'dif' subscript refers to a first-order FIR differentiator, which is a two-point numerical (backward) difference. The processing chain may incorporate any number of these units ($K_0 \geq 0$), placed before and/or after the low-pass filter. The primary purpose of these (pre or aft) units is to derive estimates of phase derivatives, with respect to time, e.g. instantaneous frequency (for $K_0 = 1$) from the low-pass filtered instantaneous phase. A secondary purpose of these (pre) units is to reduce the angular rate of change, at the low-pass filter input, thus the incidence of angle track divergence, due to angle unwrapping errors. The 'HPF' subscript refers to the combined response of all (pre and aft) differentiators in series. Thus, for

$K_0 = 0$: $H_{\text{HPF}}(z) = H_{\text{dif}}^0(z) = \left(\frac{z-1}{z}\right)^0$ and $h_{\text{HPF}}[m] = [1]$ $\qquad$ (3.15a)

$K_0 = 1$: $H_{\text{HPF}}(z) = H_{\text{dif}}^1(z) = \left(\frac{z-1}{z}\right)^1$ and $h_{\text{HPF}}[m] = \begin{bmatrix} 1 & -1 \end{bmatrix}$ $\qquad$ (3.15b)

$K_0 = 2$: $H_{\text{HPF}}(z) = H_{\text{dif}}^2(z) = \left(\frac{z-1}{z}\right)^2$ and $h_{\text{HPF}}[m] = \begin{bmatrix} 1 & -2 & 1 \end{bmatrix}$ $\qquad$ (3.15c)

$K_0 = 3: H_{\mathrm{HPF}}(z) = H_{\mathrm{dif}}^3(z) = \left(\frac{z-1}{z}\right)^3$ and $h_{\mathrm{HPF}}[m] = \begin{bmatrix} 1 & -3 & 3 & -1 \end{bmatrix}$. (3.15d)

For the *coloured-noise* case, with a single differentiator ($K_0 = 1$) and a constant frequency signal ($K_\theta = 1$), the coefficients of an FIR filter that minimizes the variance ($\sigma_\omega^2$) for a constant frequency is provided in equation (16) of Kay's 1989 paper [1]. For the more general cases considered here, i.e. for an instantaneous-phase model of $K_\theta$th order ($K_\theta > 0$), with $K_0$ first-order differentiators in series ($K_0 \geq 0$), it is readily derived using linear regression [32]. For unbiased estimates, in the absence of phase unwrapping errors, $K_1$ monomial regressors are considered, with $K_1 = K_\theta - K_0 + 1$, for a fitted polynomial with a degree (or order) of $K_1 - 1$. For minimised estimation error variance, the inverse of the coloured-noise covariance-matrix, is also considered to decorrelate (i.e. whiten) the residuals. The coefficients of this general FIR smoothing filter are therefore found by minimising the whitened sum-of-squared residuals using

$\boldsymbol{h} = \boldsymbol{x}(q)\{\boldsymbol{X}^T\boldsymbol{W}\boldsymbol{X}\}^{-1}\boldsymbol{X}^T\boldsymbol{W}$ where (3.16)

$\blacksquare^T$ is the transpose operator and $\blacksquare^{-1}$ is a matrix inverse

$\boldsymbol{h}$ is an $M \times 1$ vector with elements $h_{\mathrm{LPF}}[m]$

$\boldsymbol{X}$ is an $M \times K_1$ (Vandermonde) matrix with the regressor monomials as its columns, thus the element in the $m$th row (for $m = 0 \ldots M - 1$) and the $k_1$th column (for $k_1 = 0 \ldots K_1 - 1$) is $m^{k_1}$

$\boldsymbol{x}(q)$ is the $1 \times K_1$ 'synthesis' vector, that evaluates the locally fitted polynomial at $n - q$ or (i.e. where $m = q$) with elements $q^{k_1}$ (for $k_1 = 0 \ldots K_1 - 1$)

$\boldsymbol{W}$ is the 'whitening' matrix, with $\boldsymbol{W} = \boldsymbol{P}_{\varepsilon_\omega}^{-1}$ where $\boldsymbol{P}_{\varepsilon_\omega}^{-1}$ is the $M \times M$ (Toeplitz) covariance matrix of the coloured noise introduced by the $K_0$ differentiators.

Elements of the discrete-time coloured-noise autocorrelation function $r_{\varepsilon_\omega}[l]$, run along the $l$th off-diagonal of the covariance matrix, where $l$ is the lag index (samples), and $r_{\varepsilon_\omega}[l]$ is found via the inverse discrete-time Fourier transform of the noise PSD, i.e. $P_{\varepsilon_\omega}(\omega)$, e.g. for $K_0 = 1$, $K_0 = 2$ & $K_0 = 3$ (respectively):

$$\boldsymbol{P}_{\varepsilon_\omega} = \frac{1}{2}\begin{bmatrix} 2 & -1 & 0 & 0 & 0 & \\ -1 & 2 & -1 & 0 & 0 & \ddots \\ 0 & -1 & 2 & -1 & 0 & \ddots \\ & & \ddots & \ddots & \ddots & \end{bmatrix}$$ (3.17a)

$$\boldsymbol{P}_{\varepsilon_\omega} = \frac{1}{6}\begin{bmatrix} 6 & -4 & 1 & 0 & 0 & 0 & 0 & \\ -4 & 6 & -4 & 1 & 0 & 0 & 0 & \ddots \\ 1 & -4 & 6 & -4 & 1 & 0 & 0 & \ddots \\ 0 & 1 & -4 & 6 & -4 & 1 & 0 & \ddots \\ & & \ddots & \ddots & \ddots & \ddots & \ddots & \end{bmatrix} \text{ and }$$ (3.17b)

$$\boldsymbol{P}_{\varepsilon_\omega} = \frac{1}{20}\begin{bmatrix} 20 & -15 & 6 & -1 & 0 & 0 & 0 & 0 & 0 & \\ -15 & 20 & -15 & 6 & -1 & 0 & 0 & 0 & 0 & \ddots \\ 6 & -15 & 20 & -15 & 6 & -1 & 0 & 0 & 0 & \ddots \\ -1 & 6 & -15 & 20 & -15 & 6 & -1 & 0 & 0 & \ddots \\ 0 & -1 & 6 & -15 & 20 & -15 & 6 & -1 & 0 & \ddots \\ & & \ddots & \ddots & \ddots & \ddots & \ddots & \ddots & \ddots & \end{bmatrix}.$$ (3.17c)

In the constant-frequency case considered by Kay [1], with $K_\theta = 1$ and $K_0 = 1$, $h_{\mathrm{LPF}}[m]$ is positive for all $m$ and symmetric about its midpoint at $m = (M - 1)/2$ (for perfect phase linearity) with a quadratic taper, for lower sidelobes (which attenuates high-frequency noise) and a wider bandwidth (for improved tracking of frequencies that are only approximately constant), relative to a rectangular weight. This tapered weight applies extra attenuation, where the PSD of the noise sequence $\varepsilon_\omega$ (i.e. $P_{\varepsilon_\omega}$) is elevated. For the more general swept-frequency cases considered here, with $K_\theta > 1$, $h_{\mathrm{LPF}}[m]$ has both positive and negative values (i.e. it is somewhat oscillatory) for a wider bandwidth. Using $K_0 > 1$ further facilitates the lowering of high-frequency sidelobes and broadening of the low-frequency mainlobe.

For an unbiased frequency estimator, and in the absence of phase unwrapping errors, the expected variance of the frequency estimation error ($\hat{\omega}_\psi[n] - \omega_\psi[n]$) at steady state is evaluated analytically (or by 'analysis') from the filter response using

$$\sigma_\omega^2 = v_{\text{BPF}}/\text{SNR where} \tag{3.18}$$

$\text{SNR} = A^2/\sigma_\varepsilon^2$ and

$v_{\text{BPF}}$ is the coloured-noise gain (CNG) of the LPF, which is equivalent to the white-noise gain (WNG) of the band-pass filter (BPF) formed by cascading the LPF with the HPF. It is evaluated from the filter response using

$$v_{\text{BPF}} = \int_{-\pi}^{\pi} |H_{\text{BPF}}(e^{i\omega})|^2 \, d\omega/2\pi \text{ or from Parseval's theorem} \tag{3.19a}$$

$$v_{\text{BPF}} = \sum_{m=0}^{M-1} |h_{\text{BPF}}[m]|^2 . \tag{3.19b}$$

These BPF responses are obtained using

$$H_{\text{BPF}}(e^{i\omega}) = H_{\text{HPF}}(e^{i\omega})H_{\text{LPF}}(e^{i\omega}) \text{ and} \tag{3.20a}$$

$$h_{\text{BPF}}[m] = h_{\text{HPF}}[m] \circledast h_{\text{LPF}}[m] \text{ where} \tag{3.20b}$$

$\circledast$ is the convolution operator.

Note that these expressions also apply when the instantaneous frequency is estimated using an HPF that is applied *after* the LPF, so that the LPF operates on the unwrapped phase measurements (i.e. $\bar{\theta}$) instead of unwrapped frequency measurements (i.e. $\bar{\omega}$). For estimates of instantaneous phase, the HPF is omitted from the processing chain, i.e. $K_0 = 0$, thus $\boldsymbol{P}_{\varepsilon_\omega} = \boldsymbol{I}$ (where $\boldsymbol{I}$ is the identity matrix) and the white-noise gain (WNG) of the LPF, i.e. $v_{\text{LPF}}$, is instead used to determine the expected variance of the phase estimation error ($\hat{\theta}[n] - \theta[n]$), i.e. $\sigma_\theta^2$. The WNG is also used here to characterise the (impulse and frequency) response of low-pass estimation filters in isolation.

# 4   Recursive filter design

The discrete-time transfer function of the low-pass filter (LPF) is expressed as a linear combination of $K_\varphi$ first-order basis-functions $\varphi_k$, i.e.

$$H_{\text{LPF}}(z) = \sum_{k_\varphi=0}^{K_\varphi-1} c_{k_\varphi} \varphi_{k_\varphi}(z) \text{ where} \tag{4.1}$$

$$\varphi_{k_\varphi}(z) = \frac{z}{z - p_{k_\varphi}}$$

$p_{k_\varphi}$ is the (complex) pole of the $k_\varphi$th basis function and

$c_{k_\varphi}$ are the (complex) linear coefficients, to be determined.

Each (complex) basis function is a damped oscillator, with decreased damping (thus increased bandwidth) as $\left| p_{k_\varphi} \right| \to 0$ and decreased oscillation frequency (thus decreased bandwidth) as $\angle p_{k_\varphi} \to 0$. The poles of the $K_\varphi$ basis functions correspond to the $K_\varphi$ poles of a $K_\varphi$th-order analogue prototype. For a stable filter with approximately linear phase and approximately constant magnitude over a low-frequency passband of specified width, with low gain at high frequencies, a Bessel filter prototype, with a 3 dB cut-off frequency of $\omega_c$ ($\omega_c = 2\pi f_c$), is recommended. Its poles are discretised using

$p_{k_\varphi} = e^{\tilde{p}_{k_\varphi}}$ for $k_\varphi = 0 \dots K_\varphi - 1$ where $\qquad$ (4.2)

$\tilde{p}_{k_\varphi}$ are the continuous-time poles of the analogue prototype in the complex s-plane and

$p_{k_\varphi}$ are the discrete-time poles of the basis set in the complex z-plane.

The frequency response and impulse response of the filter in (4.1) are therefore expressed as follows:

$H_{\text{LPF}}(e^{i\omega}) = \sum_{k_\varphi=0}^{K_\varphi-1} c_{k_\varphi} \varphi_{k_\varphi}(e^{i\omega})$ and $\qquad$ (4.3a)

$h_{\text{LPF}}[m] = \sum_{k_\varphi=0}^{K_\varphi-1} c_{k_\varphi} \varphi_{k_\varphi}[m]$ where $\qquad$ (4.3b)

$\varphi_{k_\varphi}(e^{i\omega})$ is the frequency response of the $k_\varphi$th basis function and

$\varphi_{k_\varphi}[m]$ is the impulse response of the $k_\varphi$th basis function; reached via the inverse $\mathcal{Z}$-transform of $\varphi_{k_\varphi}(z)$.

The poles are set (equal to $p_{k_\varphi}$) so that the designed filter has the desired approximate bandwidth. The zeros are determined (from the $p_{k_\varphi}$ and $c_{k_\varphi}$ coefficients) to minimize the coloured noise gain and to ensure that $K_\theta$th-degree polynomials are tracked without bias at steady state. A wider bandwidth increases the noise gain but decreases the time required for the filter to reach steady state. Filters with a wider bandwidth are also able to track nearly polynomial inputs (e.g. low-frequency sinusoids) with reduced bias and quickly adjust to sudden changes in the polynomial coefficients. The linear coefficients ($c_{k_\varphi}$) and the passband group-delay ($q$) are chosen to ensure that $H_{\text{LPF}}(e^{i\omega})$ minimizes the coloured noise gain while satisfying derivative constraints at dc (where $\omega = 0$) for unbiased polynomial tracking.

The coloured noise gain of the IIR filter is determined by the substituting (4.3a) into (3.19), thus

$\upsilon_{\text{BPF}} = \boldsymbol{c}^\dagger \boldsymbol{S} \boldsymbol{c}$ where $\qquad$ (4.4)

$\blacksquare^\dagger$ is the conjugate transpose operator

$\boldsymbol{c}$ is a $K_\varphi \times 1$ vector with the $c_{k_\varphi}$ coefficients as elements, for $k_\varphi = 0 \dots K_\varphi - 1$ and

$\boldsymbol{S}$ is a $K_\varphi \times K_\varphi$ Hermitian matrix.

The elements of $\boldsymbol{S}$ are definite integrals containing the noise PSD and pairwise basis-function products

$S_{k_m,k_n} = \int_{-\pi}^{\pi} \varphi_{k_m}^*(e^{i\omega}) P_{\varepsilon_\omega}(\omega) \varphi_{k_n}(e^{i\omega}) \, d\omega/2\pi$ or with $\qquad$ (4.5a)

$P_{\varepsilon_\omega}(\omega) = \left| H_{\text{HPF}}(e^{i\omega}) \right|^2 = H_{\text{HPF}}^*(e^{i\omega}) H_{\text{HPF}}(e^{i\omega})$ we have $\qquad$ (4.5b)

$S_{k_m,k_n} = \int_{-\pi}^{\pi} \varphi_{k_m}^*(e^{i\omega}) H_{\text{HPF}}^*(e^{i\omega}) H_{\text{HPF}}(e^{i\omega}) \varphi_{k_n}(e^{i\omega}) \, d\omega/2\pi \, .$ $\qquad$ (4.5c)

For general noise distributions, these frequency-domain integrals are readily evaluated as infinite sums in the time domain (following Parseval's theorem). For IIR filters, with $M = \infty$, the element in the $k_m$th row and $k_n$th column of $\boldsymbol{S}$ is therefore

$S_{k_m,k_n} = \sum_{n=0}^{\infty} g_{k_m}^*[n] g_{k_n}[n]$ where $\qquad$ (4.6a)

$g_k[n] = h_{\text{HPF}}[n] \circledcirc \varphi_k[n] = \sum_{m=0}^{\infty} h_{\text{HPF}}[m] \varphi_k[n-m] \, .$ $\qquad$ (4.6b)

In practice, these summations are performed over a large interval that contains all but a negligible proportion of the impulse-response cross-power, for a numerical result that is a sufficient approximation of the exact analytic result.

As indicated above, $r_{\varepsilon_\omega}[l]$ is the auto-correlation vector corresponding to the PSD of the (coloured) noise to be suppressed. In the current context, the (high frequency) noise is due to the differentiators in the processing chain. In the general case ($K_0 \geq 0$), the autocorrelation vector is unity at $l = 0$; in the white noise case ($K_0 = 0$), it is zero elsewhere. The form of the PSD, i.e. $P_{\varepsilon_\omega}(\omega)$, may be appreciated by considering the Laplace transform of an ideal and unrealizable $K_0$th-order differentiation operation in continuous time: $s^{K_0}$. Evaluating this transform along the imaginary axis of the complex s-plane yields its frequency response: $(i\Omega)^{K_0}$. Thus, as the number of differentiators increases, high-frequency gain increases, and low-frequency gain decreases. The high-frequency attenuation of the low-pass filter design therefore increases with $K_0$ and (in the absence of ill-conditioning) the filter's ability to suppress the noise increases as the degrees of freedom are increased, i.e. with the size of the basis set $K_\varphi$.

Minimisation of the noise gain determines how non-signal inputs (e.g. noise and interference) are treated while derivative constraints at $\omega = 0$ (i.e. dc) determine how signal inputs (i.e. the phase polynomial) are treated. Degrees of freedom are therefore 'consumed' to ensure that $K_1$ derivative constraints are satisfied, for unbiased steady-state tracking of a polynomial input with a degree of $K_1 - 1$. These constraints on the low-pass filter's frequency response are stated as follows:

$$\left\{\frac{d^{k_1}}{d\omega^{k_1}} H_{\mathrm{LPF}}(e^{i\omega})\right\}\Big|_{\omega=0} = (-iq)^{k_1} \text{ for } k_1 = 0 \dots K_1 - 1 \tag{4.7}$$

or after substitution of (4.3a) into (4.7)

$$\sum_{k_\varphi=0}^{K_\varphi-1} c_{k_\varphi} \left\{\frac{d^{k_1}}{d\omega^{k_1}} \varphi_{k_\varphi}(e^{i\omega})\right\}\Big|_{\omega=0} = (-iq)^{k_1} \text{ for } k_1 = 0 \dots K_1 - 1 \ . \tag{4.8}$$

The basis set must be constructed so that the degrees of freedom are greater than or equal to the number of constraints (i.e. $K_\varphi \geq K_1$). 'Surplus' degrees of freedom (i.e. $K_\varphi - K_1$) are available for the minimization of noise gain. Note that the $K_0$ parameter does not consume degrees of freedom; rather, it is used to define the shape of the noise PSD, thus the gain profile of the low-pass filter, away from the dc limit.

Reasonable filtering solutions are not guaranteed for all parameter combinations. For instance, ill conditioning and rounding errors may yield coefficients with non-negligible imaginary components, particularly when an unsuitable basis set is used. Furthermore, using $K_0 > 1$ is recommended for a smooth impulse response (near $m = 0$) with good high-frequency attenuation. When using $K_0 \leq 1$, an additional derivative constraint at $\omega = \pi$ (i.e. the Nyquist frequency) may be imposed to encourage a low-pass solution [27], [29] & [33], i.e.

$$\left\{\frac{d}{d\omega} H_{\mathrm{LPF}}(e^{i\omega})\right\}\Big|_{\omega=\pi} = 0 \tag{4.9}$$

although, this option is not considered further here.

The linear coefficients that minimize (4.4) subject to the equality constraints in (4.8) are found via the method of Lagrange multipliers using

$$\begin{bmatrix} \mathbf{0}_{K_\varphi \times 1} \\ \boldsymbol{d}(q)_{K_1 \times 1} \end{bmatrix} = \begin{bmatrix} \boldsymbol{S}_{K_\varphi \times K_\varphi} & \boldsymbol{\Phi}^\dagger_{K_\varphi \times K_1} \\ \boldsymbol{\Phi}_{K_1 \times K_\varphi} & \mathbf{0}_{K_1 \times K_1} \end{bmatrix} \begin{bmatrix} \boldsymbol{c}_{K_\varphi \times 1} \\ \boldsymbol{\lambda}_{K_1 \times 1} \end{bmatrix} \text{ where} \tag{4.10}$$

$\boldsymbol{\lambda}$ is a $K_1 \times 1$ vector containing the Lagrange multipliers (which are ignored)

$\boldsymbol{c}$ is a $K_\varphi \times 1$ vector; the $k_\varphi$th element is $c_{k_\varphi}$

$\boldsymbol{S}$ is the $K_\varphi \times K_\varphi$ matrix of (numerically computed) definite integrals defined in (4.5)

$\boldsymbol{\Phi}$ is a $K_1 \times K_\varphi$ matrix containing the dc derivatives of the basis functions; the element in the $k_1$th row and $k_\varphi$th column is

$$\Phi_{k_1, k_\varphi} = \left\{ \frac{d^{k_1}}{d\omega^{k_1}} \varphi_{k_\varphi}(e^{i\omega}) \right\} \bigg|_{\omega=0}$$

$\boldsymbol{d}(q)$ is a $K_1 \times 1$ vector containing the desired dc derivatives; the element in the $k_1$th row is

$$d_{k_1} = (-iq)^{k_1} \ .$$

As discussed in [32], equations in the form of (4.10) may be solved using

$$\boldsymbol{c} = \boldsymbol{d}^\dagger(q) \left\{ \boldsymbol{\Phi S}^{-1} \boldsymbol{\Phi}^\dagger \right\}^{-1} \boldsymbol{\Phi S}^{-1} \tag{4.11}$$

which has the same form as the time-domain regression formulation, for the FIR filters in (3.16), with $\boldsymbol{\Phi} = \boldsymbol{X}^T$ and $\boldsymbol{S}^{-1} = \boldsymbol{W}$. Indeed, the FIR filters are obtained using this frequency-domain procedure for IIR filters when basis functions with repeated poles at the origin of the complex z-plane are used [29] & [32], i.e.

$$\varphi_{k_\varphi}(z) = \frac{1}{z^{k_\varphi}} \ . \tag{4.12}$$

Furthermore, Laguerre-type filters are obtained when real repeated poles at $z = p$ on the $(0,1)$ interval are used [33], i.e.

$$\varphi_{k_\varphi}(z) = \frac{1}{(z-p)^{k_\varphi}} \ . \tag{4.13}$$

For specified $q$, e.g. using $q = -1$ for the one-sample lead of the predictor, the solution is readily found using (4.11). However, if the passband group-delay is unspecified, e.g. for the estimator, then $q$ may be used as a free parameter and set to minimize the CNG of the filter, which is evaluated using

$$v_{\text{BPF}}(q) = \boldsymbol{d}(q)^\dagger \left\{ \boldsymbol{\Phi S}^{-1} \boldsymbol{\Phi}^\dagger \right\}^{-1} \boldsymbol{d}(q) \ . \tag{4.14}$$

The CNG of the filter is a polynomial in $q$ thus the $2(K_1 - 1) - 1$ stationary points (i.e. local minima and maxima) are found by finding the roots of the $v'_{\text{BPF}}(q)$ polynomial, where

$$v'_{\text{BPF}}(q) = dv_{\text{BPF}}(q)/dq \ . \tag{4.15}$$

For $K_1 = 1$, the solution is independent of $q$ thus the passband phase of these filters cannot be shifted, i.e. forward for the predictor or backward for the estimator. Passband phase-control is obtained for $K_1 = 2$, and the single solution minimizes the variance of the estimator. Of the three solutions in the $K_1 = 3$ case, the solutions with the smallest and largest delay have approximately the same CNG (i.e. $v_{\text{BPF}}$); whereas the middle solution has the largest CNG but the best phase linearity and this optimal value of $q$ was used by all IIR estimators with $K_1 = 3$ in Section 6. For $K_1 > 3$, the minimum-phase solution with the lowest group delay is reached using the smallest value of $q$. Alternatively, the solution with smallest CNG may be preferred. The selected $q$ solution is then substituted into $\boldsymbol{d}(q)$, then (4.11) is solved for $\boldsymbol{c}$.

## 5  Recursive filter realization

The coefficients of the linear difference equation for the discrete-time realization of the digital filter are obtained from $p_{k_\varphi}$ and $c_{k_\varphi}$ by placing the summation terms in (4.1) over a common denominator $A(z)$ then collecting numerator terms to form the polynomial $B(z)$ such that

$$H_{\text{LPF}}(z) = B(z)/A(z) \ . \tag{5.1}$$

Thus $p_{k_\varphi}$ are combined to form the coefficients of the denominator polynomial $A(z)$, and the poles of the filter are equal to the poles of the basis set (for $k_\varphi = 0 \ldots K_\varphi - 1$); whereas $p_{k_\varphi}$ and $c_{k_\varphi}$ are combined to form the coefficients of the numerator polynomial $B(z)$, thus the zeros of the filter.

The discrete-time transfer function $H_{\mathrm{LPF}}(z)$, defines the relationship between the filter input sequence $x[n]$ with $\mathcal{Z}$-transform $X(z)$, and the filter output sequence $y[n]$ with $\mathcal{Z}$-transform $Y(z)$, i.e.

$$H_{\mathrm{LPF}}(z) = Y(z)/X(z) \,. \tag{5.2}$$

Substitution of (5.2) into (5.1),

$$A(z)Y(z) = B(z)X(z) \tag{5.3}$$

taking the inverse $\mathcal{Z}$-transform of both sides, then re-arranging, yields the filter's $K_\varphi$th-order linear difference equation

$$y[n] = \sum_{k_\varphi=0}^{K_\varphi-1} b[k_\varphi]x[n-k_\varphi] - \sum_{k_\varphi=1}^{K_\varphi} a[k_\varphi]y[n-k_\varphi] \text{ where} \tag{5.4}$$

$b[k_\varphi]$ is the $k_\varphi$th coefficient of the $B(z)$ polynomial, with $b[K_\varphi] = 0$ and

$a[k_\varphi]$ is the $k_\varphi$th coefficient of the $A(z)$ polynomial, with $a[0] = 1$.

The $K_\varphi$th-order IIR filters for estimation and prediction are realized using the linear state-space (LSS) recursion

$$\boldsymbol{w}[n] = \boldsymbol{G}\boldsymbol{w}[n-1] + \boldsymbol{H}x[n] \text{ and} \tag{5.5a}$$

$$y[n] = \boldsymbol{C}\boldsymbol{w}[n] \text{ with} \tag{5.5b}$$

$$\boldsymbol{G} = \begin{bmatrix} -a[1] & -a[2] & \cdots & -a[K_\varphi-2] & -a[K_\varphi-1] & -a[K_\varphi-0] \\ 1 & 0 & \cdots & 0 & 0 & 0 \\ 0 & 1 & \cdots & 0 & 0 & 0 \\ \vdots & \vdots & \ddots & \vdots & \vdots & \vdots \\ 0 & 0 & \cdots & 1 & 0 & 0 \\ 0 & 0 & \cdots & 0 & 1 & 0 \end{bmatrix}_{K_\varphi \times K_\varphi}, \boldsymbol{H} = \begin{bmatrix} 1 \\ 0 \\ \vdots \\ 0 \\ 0 \\ 0 \end{bmatrix}_{K_\varphi \times 1}$$

$$\boldsymbol{C} = \begin{bmatrix} b[0] & b[1] & \cdots & b[K_\varphi-3] & b[K_\varphi-2] & b[K_\varphi-1] \end{bmatrix}_{1 \times K_\varphi} \text{ and}$$

$\boldsymbol{w}[n]$ is the $K_\varphi \times 1$ internal state vector at the time of the $n$th input sample. Note that in this so-called "canonical" form, with arithmetic operations minimized (due to the sparsity of $\boldsymbol{H}$ & $\boldsymbol{G}$), the elements of this state vector do not correspond to derivative estimates, thus they are simply delayed accumulator registers with no physical interpretation. Although, the coordinates of this system may be linearly transformed so that that the internal states are more meaningful, if desired [27].

The estimator (for optimal smoothing) and predictor (for phase unwrapping) have the same poles thus their $\boldsymbol{G}$ matrices are identical, and a common input and feedback section may be used for both, i.e. (5.5a); however, they have different zeros thus different $\boldsymbol{C}$ vectors are used in the output sections, i.e. in (5.5b). The estimator and the predictor recursions are combined as follows:

$$\bar{x}[n] = \hat{x}[n-1] + \angle e^{i(\bar{x}[n]-\hat{x}[n-1])} \tag{5.6a}$$

$$\boldsymbol{w}[n] = \boldsymbol{G}\boldsymbol{w}[n-1] + \boldsymbol{H}\bar{x}[n] \tag{5.6b}$$

$$\hat{y}[n] = \boldsymbol{C}_{\mathrm{est}}\boldsymbol{w}[n] \text{ and} \tag{5.6c}$$

$$\hat{x}[n] = \boldsymbol{C}_{\mathrm{prd}}\boldsymbol{w}[n] \text{ where} \tag{5.6d}$$

$C_{\text{est}}$ contains the $b$ coefficients of the estimator and

$C_{\text{prd}}$ contains the $b$ coefficients of the predictor.

The input is either the raw instantaneous phase or the raw instantaneous frequency, i.e. $\tilde{x}[n] = \tilde{\theta}[n]$ or $\tilde{x}[n] = \tilde{\omega}[n]$. The input to the feedback section is the corresponding unwrapped value, i.e. $\bar{x}[n] = \bar{\theta}[n]$ or $\bar{x}[n] = \bar{\omega}[n]$. The output of the estimator is the corresponding estimated value, i.e. $\hat{y}[n] = \hat{\theta}[n-q]$ or $\hat{y}[n] = \hat{\omega}[n-q]$. The output of the predictor is the predicted unwrapped input at the time of the next complex sample, i.e. $\hat{x}[n] = \bar{\theta}[n+1]$ or $\hat{x}[n] = \bar{\omega}[n+1]$. The recursion is initialized using

$$\hat{x}[0] = \tilde{x}[0] \text{ and} \tag{5.7a}$$

$$\boldsymbol{w}[0] = \boldsymbol{w}_0[0]\tilde{x}[0] \tag{5.7b}$$

where $\boldsymbol{w}_0$ is the $K_\varphi \times 1$ state initialization vector, which is the steady-state response of the system to a unit step input. It may be computed analytically in the complex z-plane via the final-value theorem, or in the discrete-time domain via a loop until convergence. This logic initializes the internal states of the filter with the final values that would be observed if the initial input had been applied for an infinite time.

# 6    Simulation and analysis

Monte Carlo (MC) simulations were performed to analyse the behaviour of the proposed digital filters for instantaneous phase and frequency estimation. Three signal types were generated, each with one thousand samples (i.e. $N = 1000$). Three system 'configurations' were considered for each signal 'type', for nine simulation 'scenarios'.

The system is comprised of two low-pass filters (a predictor and an estimator) operating in tandem, preceded by $\alpha$ two-point differentiators ($\alpha \geq 0$) followed by $\beta$ two-point differentiators ($\beta \geq 0$) thus $K_0 = \alpha + \beta$, where $\alpha$ and $\beta$ are the number of pre and aft differentiators, respectively. The system configurations are enumerated using the $\alpha/\beta$ doublet, and a variety of different low-pass (FIR & IIR) filters were considered for each configuration.

The received signal is a complex exponential with polynomial phase ($K_\theta \leq 3$) and it is defined by specifying its instantaneous frequency (cycles per sample) at $n = 0$, $n = (N-1)/2$ and $n = N-1$, using $f_0$, $f_1$ and $f_2$, respectively. The signal types are denoted using the $f_0/f_1/f_2$ triplet. In each MC instantiation, the phase offset (i.e. $\phi$) was pseudo-randomly generated using a uniform distribution over the $(0,2\pi)$ interval.

For a given signal type, system configuration, and set of low-pass filters, the variance of the instantaneous phase or instantaneous frequency estimate (i.e. $\sigma_\theta^2$ or $\sigma_\omega^2$) was evaluated over a range of SNR values (e.g. 0 to 20 dB in 1 dB steps), using one thousand random MC instantiations for each SNR. The sampled waveform was generated for a complex signal of unity magnitude with Gaussian-distributed complex noise added, i.e. $\text{Re}\{\varepsilon_\psi\} \sim \mathcal{N}(0, \sigma_\varepsilon^2)$ and $\text{Im}\{\varepsilon_\psi\} \sim \mathcal{N}(0, \sigma_\varepsilon^2)$, where $\text{Re}\{\varepsilon_\psi\}$ and $\text{Im}\{\varepsilon_\psi\}$ extract the real and imaginary parts of a complex number, and $\mathcal{N}(\mu, \sigma^2)$ is a normal distribution with a mean of zero and a variance of $\sigma^2$. Simulation results are presented using estimator variance (on a 10 x $\log_{10}$ scale, i.e. dB) versus SNR (dB) plots. The estimator variance observed in the simulations and the estimator variance derived from response analysis using (3.18) are plotted. These quantities are denoted using the 'sim' and 'ana' qualifiers in the legends of the plots. The number of differentiators used for filter design (denoted above using $K_0$) does not always equal the number of differentiators used in the simulation (denoted below using $\tilde{K}_0$). The CNG of each filter ($\tilde{v}_{\text{BPF}}$) is therefore computed for the simulated system configurations ($\widetilde{K}_0$) in which it is applied and provided in the tables. Note that the analysis error variance is for an *unbiased* estimator. In theory, bias may be introduced through a variety of mechanisms, such as: 1) the filter having an insufficient number of dc flatness constraints (i.e. $K_1$) for a given polynomial order (i.e. $K_\theta$) and number of differentiators (i.e. $K_0$); 2) insufficient time for the filter to reach steady state, which may take a while for narrowband filters; and 3) phase unwrapping errors. In practice, any number of other factors may also lead to a bias, such as: 1) non-polynomial phase progression; 2) correlated measurement errors at low frequency; 3) high-frequency interference; 4) amplitude modulated signals; etc.

All low-pass filters were initialized to minimize the impact of startup transients. For a given IIR filter, this is done by determining its internal state at the $n \to \infty$ limit, for a unit step input, using a loop over $n$ until changes to the internal state vector are less than a specified tolerance. For an FIR filter, the loop is truncated early when the steady-state limit is reached at $n = M - 1$. The internal state of the (FIR or IIR) filter is then initialized by multiplying the determined state-vector, by the first input sample, as indicated in (5.7). However, this does not eliminate start-up transients completely, especially for filters with a narrow bandwidth, thus only samples from $n = (N-1)/8 \dots (N-1)$ were considered in the error variance calculations.

The latency of the filters was considered when the error variance was computed. This was done by advancing the delayed output of the estimator, by $q$ samples, so it is aligned with the signal input. The high-pass FIR differentiator (with two coefficients) has a group delay of half a sample ($q = 0.5$). All FIR filters (estimators and predictors) were designed with $M = 64$, thus $q = (M-1)/2 = 31.5$ for the linear-phase estimators used here. The selected optimal value of $q$ for the IIR filters is provided in the tables.

The impulse responses and frequency responses of the low-pass filters were used to account for the trends observed in the simulations. Frequency responses (of selected filters) are represented using plots with four panels in a 2 x 2 configuration. The all-band magnitude ('mag') response, i.e. $\left|H_{\text{LPF}}\left(e^{i\omega}\right)\right|^2$ over $f = \omega/2\pi = 0.0 \dots 0.5$ is plotted in the left panels, with a linear scale in the top-left panel (to reveal the passband) and a log scale in the bottom left panel (to reveal the stopband); the passband phase-response is plotted over $f = \omega/2\pi = 0.0 \dots 0.95 f_c$ in the right panels. The phase response, i.e. $\angle H_{\text{LPF}}\left(e^{i\omega}\right)$ in degrees, is shown in the top right panel. For each filter, the phase response of an ideal linear-phase filter with a group delay of $q$ is shown using a dashed line. The deviation from phase linearity over the passband, i.e. $\angle H_{\text{LPF}}\left(e^{i\omega}\right)e^{iq\omega}$ in degrees, is plotted in the bottom right panel. Note that the terms 'passband' and 'stopband' are used loosely here, because these filters are not designed using passband and stopband specifications, although these 'band' qualifiers are still broadly applicable.

For FIR filters, the nominal bandwidth of the frequency response is $f_c = 1/M$; and for IIR filters, the nominal timescale of the impulse response is $M = 1/f_c$. All FIR filters in this section were designed using $M = 64$, which was considered to offer a reasonable compromise between estimation performance (i.e. noise reduction at high SNR) and computational complexity. All IIR filters were designed using a basis set with $K_\varphi = 5$. In the general case, non-recursive FIR filters require $M$ delay-multiply-add operations, as shown in (3.12), whereas recursive IIR filters require only $2K_\varphi$, as shown in (5.4), thus the computational complexity of the FIR filters is around 6x greater than the IIR filters.

## 6.1 Signal type 1: Constant frequency

A signal with a constant (low) frequency (i.e. $K_\theta = 1$) was processed, with $f_0 = f_1 = f_2 = 0.01$. The parameters of the low-pass FIR filters used for each system configuration are provided in Table 1. The impulse responses of the corresponding estimator and predictor are plotted in Figure 1 and Figure 2, respectively. Note that the estimators for Filters A1-D1 (all with $K_1 = 1$ and $K_0 = 0, 1, 2 \, \& \, 3$) are identical to the corresponding Filters E1-H1 (all with $K_1 = 2$ and $K_0 = 0, 1, 2 \, \& \, 3$) due to even symmetry (see Figure 1). Note also that the phase cannot be shifted for Filters A1-D1, thus their estimators and predictors are identical (see Figure 1 and Figure 2).

*Table 1. FIR filters used for signal type 1.*

| LPF ID | LPF Type | $K_1$ | $K_0$ | $M$ | $f_c$ | $q$ | $v_{\mathrm{LPF}}$ | $v_{\mathrm{BPF}}$ | $\tilde{v}_{\mathrm{BPF}}$ ($\tilde{K}_0 = 1$) | $\tilde{v}_{\mathrm{BPF}}$ ($\tilde{K}_0 = 0$) |
|---|---|---|---|---|---|---|---|---|---|---|
| A1 | FIR | 1 | 0 | 64.00 | 0.0156 | 31.500 | 0.0156 | 1.563E-02 | 4.883E-04 | 1.563E-02 |
| B1 | FIR | 1 | 1 | 64.00 | 0.0156 | 31.500 | 0.0185 | 4.371E-05 | 4.371E-05 | 1.847E-02 |
| C1 | FIR | 1 | 2 | 64.00 | 0.0156 | 31.500 | 0.0217 | 5.756E-07 | 5.977E-05 | 2.167E-02 |
| D1 | FIR | 1 | 3 | 64.00 | 0.0156 | 31.500 | 0.0244 | 1.668E-08 | 8.496E-05 | 2.439E-02 |
| E1 | FIR | 2 | 0 | 64.00 | 0.0156 | 31.500 | 0.0156 | 1.563E-02 | 4.883E-04 | 1.563E-02 |
| F1 | FIR | 2 | 1 | 64.00 | 0.0156 | 31.500 | 0.0185 | 4.371E-05 | 4.371E-05 | 1.847E-02 |
| G1 | FIR | 2 | 2 | 64.00 | 0.0156 | 31.500 | 0.0217 | 5.756E-07 | 5.977E-05 | 2.167E-02 |
| H1 | FIR | 2 | 3 | 64.00 | 0.0156 | 31.500 | 0.0244 | 1.668E-08 | 8.496E-05 | 2.439E-02 |

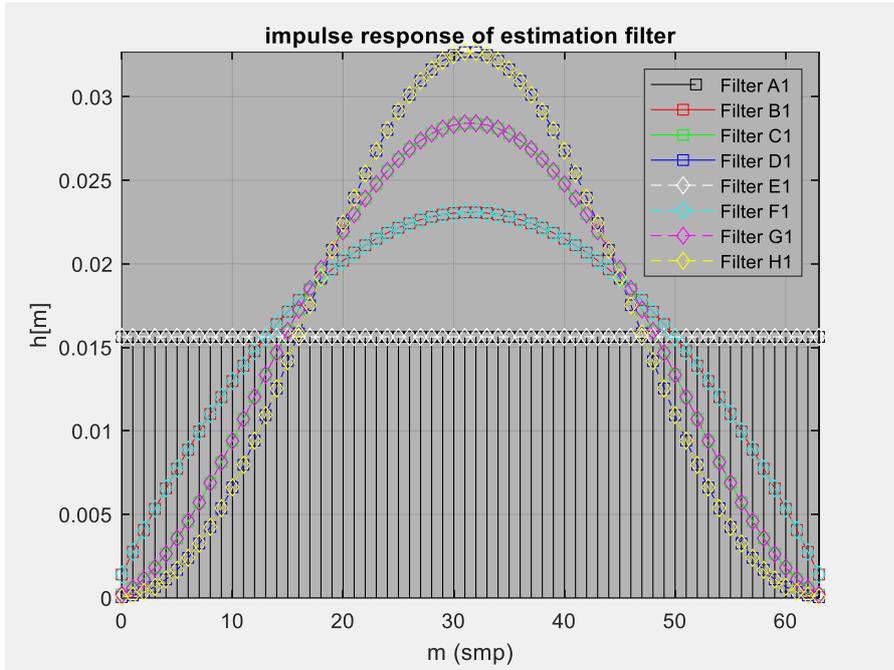

*Figure 1. Impulse responses of FIR estimators in Table 1.*

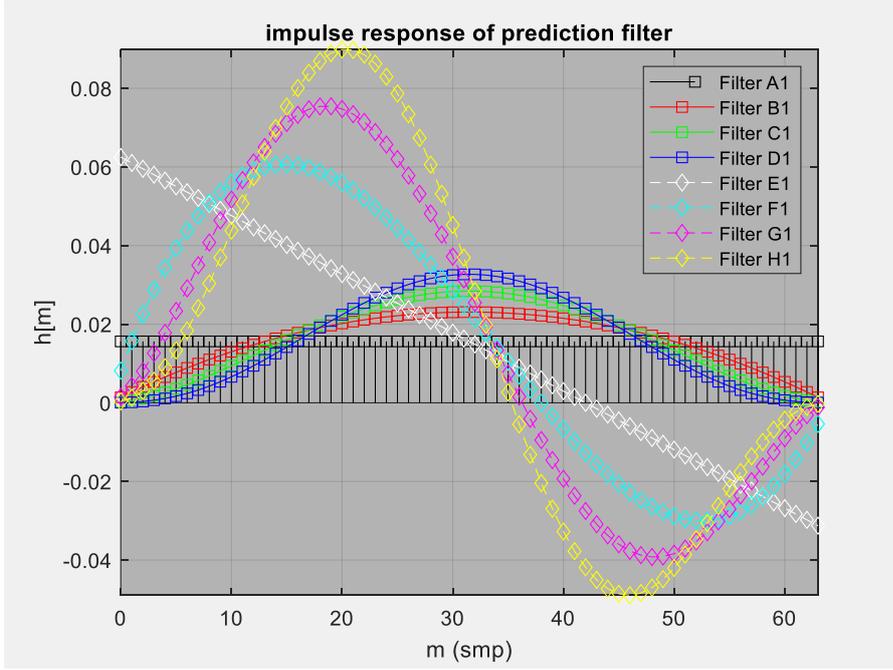

*Figure 2. Impulse responses of FIR predictors in Table 1.*

### 6.1.1 System configuration 1: Pre-differentiator

In this scenario, the instantaneous frequency of a constant-frequency signal is estimated using a differentiator at the start of the processing chain ($\alpha = 1$ & $\beta = 0$) so that the input to the low-pass filter is a constant (in the absence of noise). The complex waveform output by the analogue-to-digital converter (ADC) and the angle (i.e. frequency) measurement inputs to the low-pass filter (LPF) are plotted in the upper and lower subplots (respectively) of Figure 3. Figure 4 shows that the results computed from the simulations results agree with the results expected from analysis, at high SNR. As expected from the theory in Section 3, and as quantified by the $\tilde{v}_{\mathrm{BPF}}$ parameter in Table 1, the error variance is greatest for Filters A1 & E1 (both with $K_0 = 0$) and least for Filters B1 & F1 (both with $K_0 = 1$) that are matched to the number of differentiators used in the simulation ($\tilde{K}_0 = \alpha + \beta = 1$). All frequency estimators are unbiased (at high SNR), because $K_1 \geq K_\theta - \tilde{K}_0 + 1$, with $K_\theta = 1$ & $\tilde{K}_0 = 1$ in this scenario (Filters B1 & F1 reach the CRLB); however, at around 10 dB, all filters start to fail due to noise-induced unwrapping errors. Similar results were observed when the signal frequency was increased to $f_\psi = 0.5$, which suggests that the failure threshold is independent of signal frequency in this scenario.

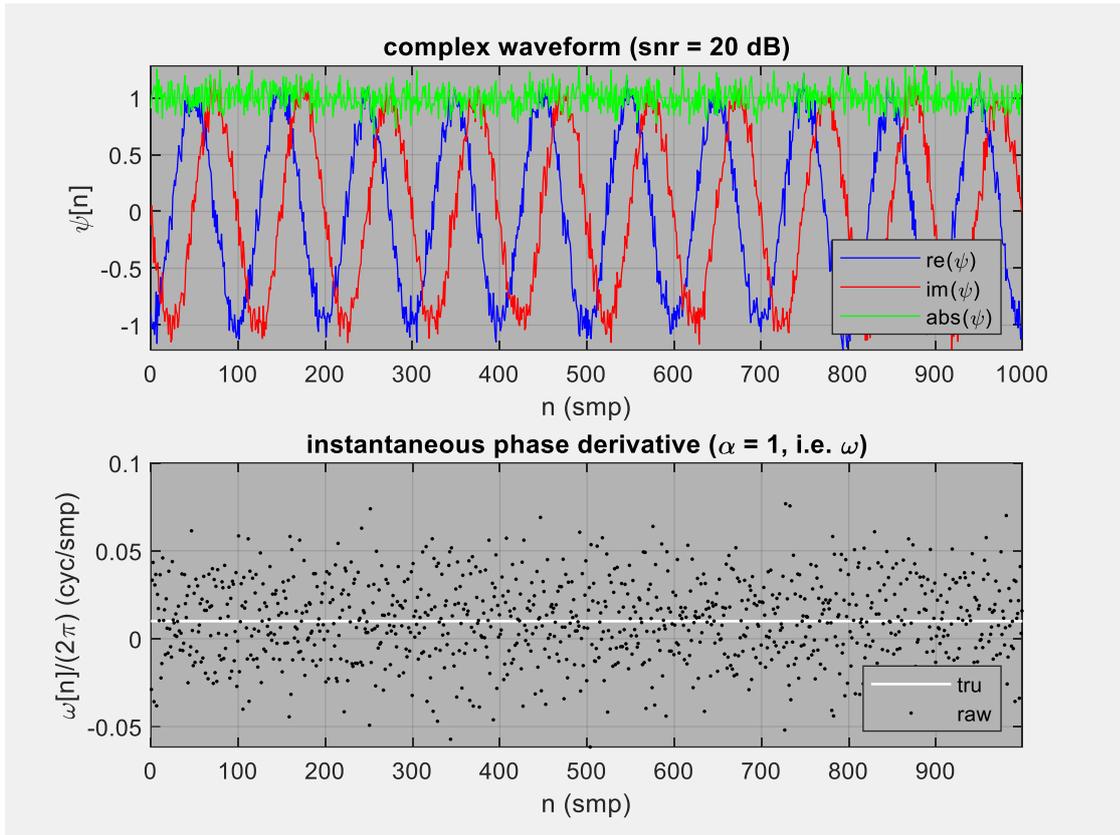

*Figure 3. Example ADC output (top) and LPF input (bottom) for signal type 1 and system configuration 1.*

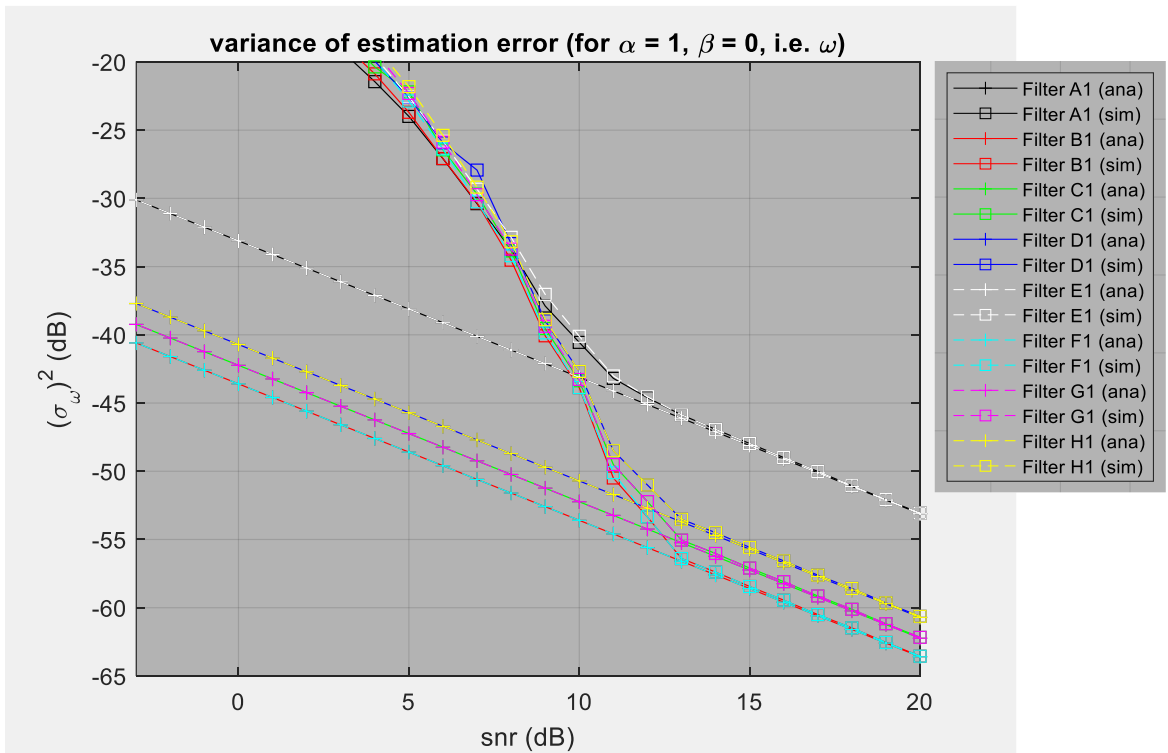

*Figure 4. Error variance versus SNR for signal type 1 and system configuration 1 using the FIR filters in Table 1.*

### 6.1.2 System configuration 2: Aft-differentiator

In this scenario, the instantaneous frequency is again estimated; however, the differentiator is placed at the end of the processing chain in this configuration ($\alpha = 0$ & $\beta = 1$), thus the low-pass filter must handle an angle input with a non-zero rate of change (see lower subplot of Figure 5 ), with noise-induced *and* signal-induced angle-wrapping events. Figure 6 shows that moving the differentiator (from 'pre' to 'aft') does not change the analysis error variance at high SNR; however, Filters E1-H1 (all with $K_1 = 2$) are accurate down to a much lower SNR because the predictors, with a one-sample lead ($q = -1$), deal with less (high-frequency) noise when making phase unwrapping decisions. However, Filters A1-D1 (all with $K_1 = 1$) fail at around 10 dB because the lagged output of the predictor ($q = 31.5$) causes phase unwrapping errors, thus divergent (i.e. biased) estimates. When the frequency of the signal is increased fivefold (to $f_\psi = 0.05$), all filters fail at a much higher SNR and Filter E1 performs the best with a threshold around 10 dB. The threshold continues to rise (i.e. unwrapping performance deteriorates) as the frequency is further increased.

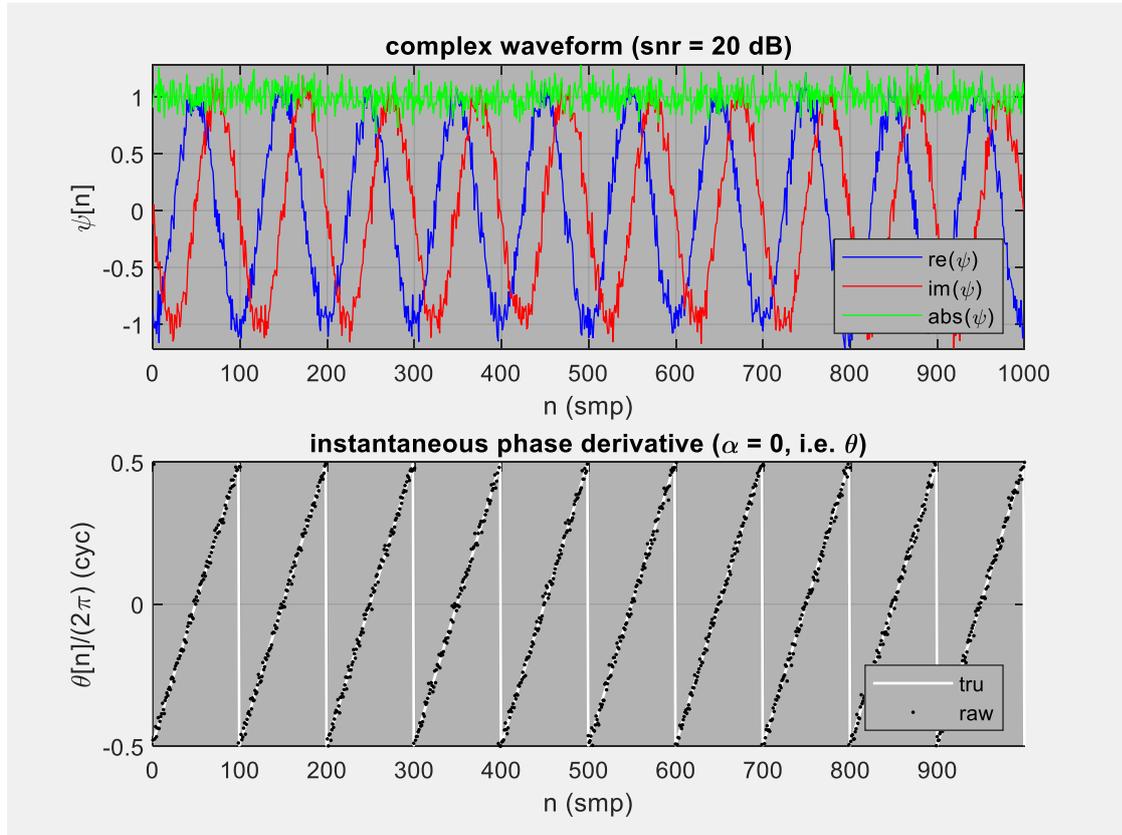

*Figure 5. Example ADC output (top) and LPF input (bottom) for signal type 1 and system configurations 2 & 3.*

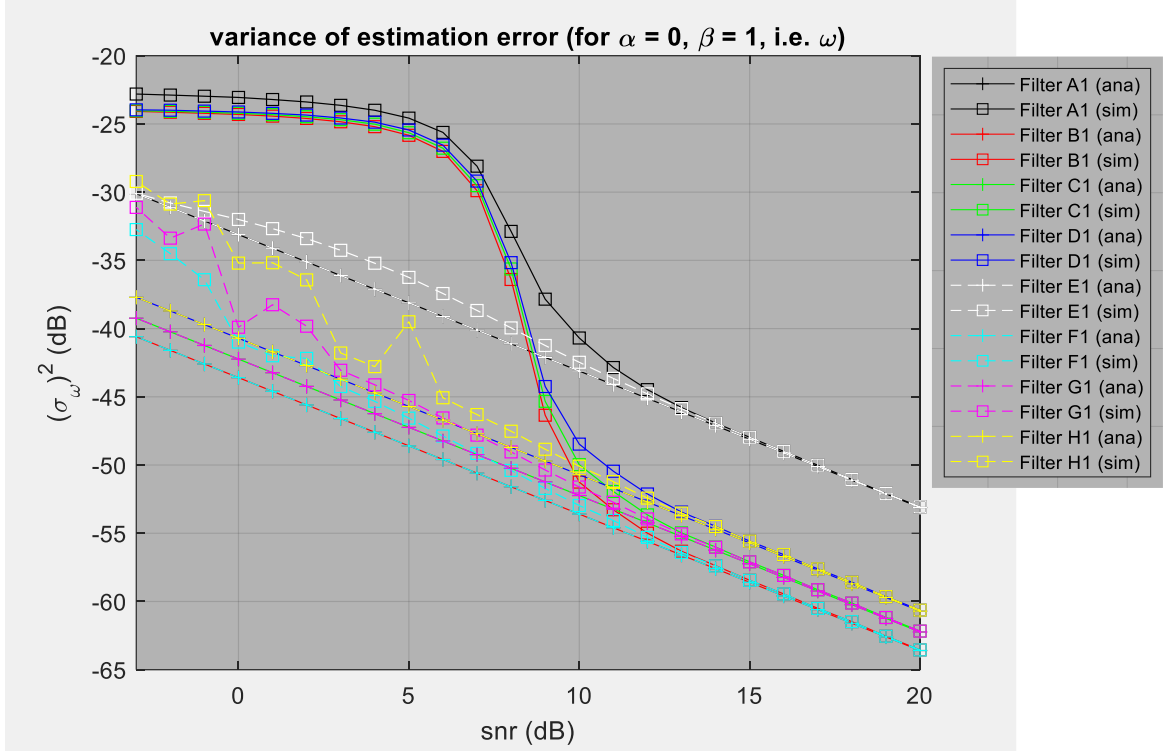

*Figure 6. Error variance versus SNR for signal type 1 and system configuration 2 using the FIR filters in Table 1.*

### 6.1.3 System configuration 3: Nil-differentiator

In this configuration, the instantaneous phase is estimated (instead of frequency), so no differentiators are used ($\alpha = 0$ & $\beta = 0$). The low-pass filter must again handle an angle input with a non-zero rate of change (see lower subplot of Figure 5 ). In theory, for this system configuration (with $\widetilde{K}_0 = \alpha + \beta = 0$) the error variance (see Figure 7) should be minimised (and the CRLB reached) for Filters A1 & E1 (both with $K_0 = \widetilde{K}_0 = 0$), because the parameters of these filters (with identical impulse responses) match the parameters of the simulated scenario. The other filters should all have a higher error variance because $K_0 > \widetilde{K}_0$. At high SNR the simulated results agree with the analysis results and as was noted for system configuration 2, Filters A1-D1 (all with $K_1 = 1$, for predictors with a phase lag) fail at around 10 dB due to phase unwrapping errors, whereas the error variance of Filters E1-H1 (all with $K_1 = 2$, for predictors with a phase lead) increases slowly down to 0 dB. As was also the case for configuration 2, all filters fail at a higher SNR when processing signals of a higher frequency. For $f_\psi = 0.05$, Filter E1 again performs the best with the threshold remaining around 10 dB.

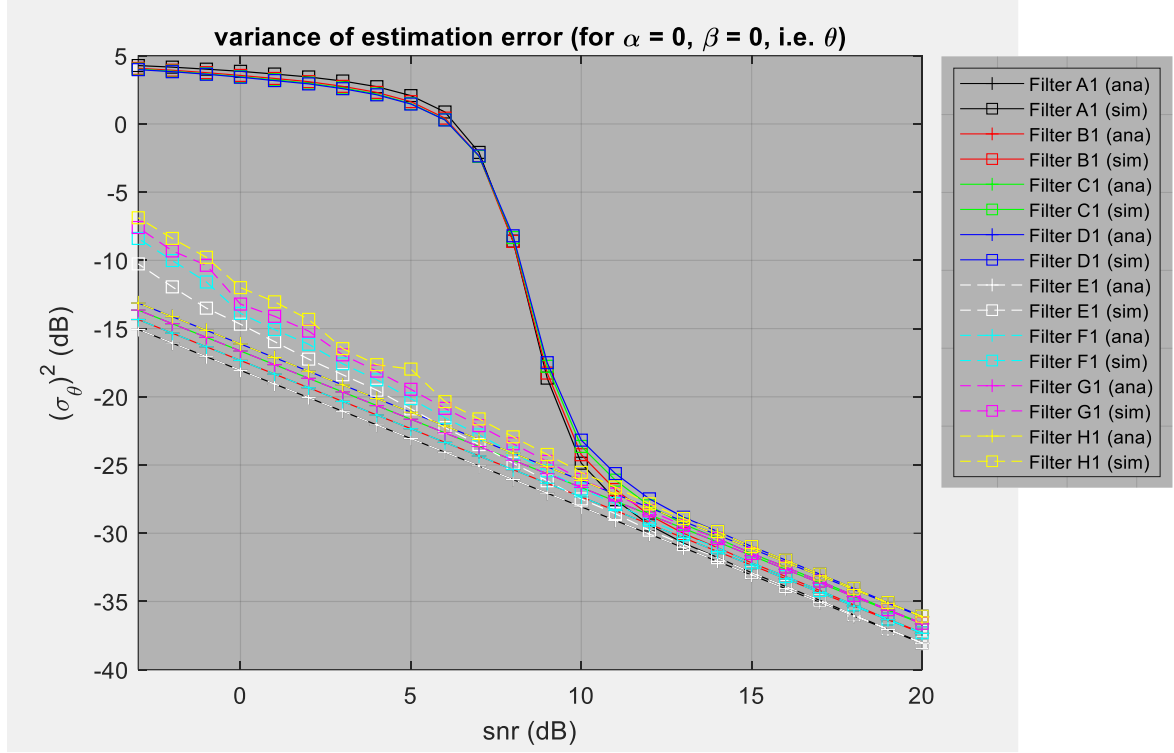

*Figure 7. Error variance versus SNR for signal type 1 and system configuration 3 using the FIR filters in Table 1.*

## 6.2  Signal type 2: Linear frequency sweep

A signal with a linear frequency sweep (i.e. $K_\theta = 2$) with $f_0 = 0.000$, $f_1 = 0.125$ & $f_2 = 0.250$ is considered here. The parameters of the FIR and IIR low-pass filters to process this waveform are provided in Table 2, Table 3 & Table 4. The filters in Table 2 are a mix of FIR and IIR filters; the two IIR filters (with $K_1 = 2$ & $K_0 = 3$ and $K_1 = 3$ & $K_0 = 3$) were designed using $f_c = 1/M$. The impulse responses of these FIR and IIR estimators are plotted in Figure 8, the corresponding impulse responses of the predictors are plotted in Figure 9. The IIR filters in Table 3 (all with $K_1 = 2$ & $K_0 = 3$) and Table 4  (all with $K_1 = 3$ & $K_0 = 3$) were designed using an extended range of bandwidths: using $f_c = (0.7, 0.8, 0.9, 1.0, 1.1, 1.2$ & $1.3)/M$ for Filters A3-G3 in Table 3 (see Figure 12 & Figure 13 for their estimator and predictor impulse responses) and $f_c = (0.5, 0.6, 0.7, 0.8, 0.9, 1.0$ & $1.1)/M$ for Filters A4-G4 in Table 4 (see Figure 14 & Figure 15 for their estimator and predictor impulse responses).

*Table 2. FIR and IIR filters used for signal type 2.*

| LPF ID | LPF Type | $K_1$ | $K_0$ | $M$ | $f_c$ | $q$ | $v_{\text{LPF}}$ | $v_{\text{BPF}}$ | $\tilde{v}_{\text{BPF}}$ ($\tilde{K}_0 = 1$) | $\tilde{v}_{\text{BPF}}$ ($\tilde{K}_0 = 0$) |
|---|---|---|---|---|---|---|---|---|---|---|
| A2 | FIR | 2 | 0 | 64.00 | 0.0156 | 31.500 | 0.0156 | 1.563E-02 | 4.883E-04 | 1.563E-02 |
| B2 | FIR | 2 | 1 | 64.00 | 0.0156 | 31.500 | 0.0185 | 4.371E-05 | 4.371E-05 | 1.847E-02 |
| C2 | FIR | 2 | 2 | 64.00 | 0.0156 | 31.500 | 0.0217 | 5.756E-07 | 5.977E-05 | 2.167E-02 |
| D2 | IIR | 2 | 3 | 64.00 | 0.0156 | 39.626 | 0.0194 | 3.644E-09 | 4.306E-05 | 1.939E-02 |
| E2 | FIR | 3 | 0 | 64.00 | 0.0156 | 31.500 | 0.0352 | 3.517E-02 | 1.212E-03 | 3.517E-02 |
| F2 | FIR | 3 | 1 | 64.00 | 0.0156 | 31.500 | 0.0385 | 2.734E-04 | 2.734E-04 | 3.851E-02 |
| G2 | FIR | 3 | 2 | 64.00 | 0.0156 | 31.500 | 0.0427 | 7.061E-06 | 3.334E-04 | 4.272E-02 |
| H2 | IIR | 3 | 3 | 64.00 | 0.0156 | 39.626 | 0.0311 | 2.149E-07 | 1.445E-04 | 3.115E-02 |

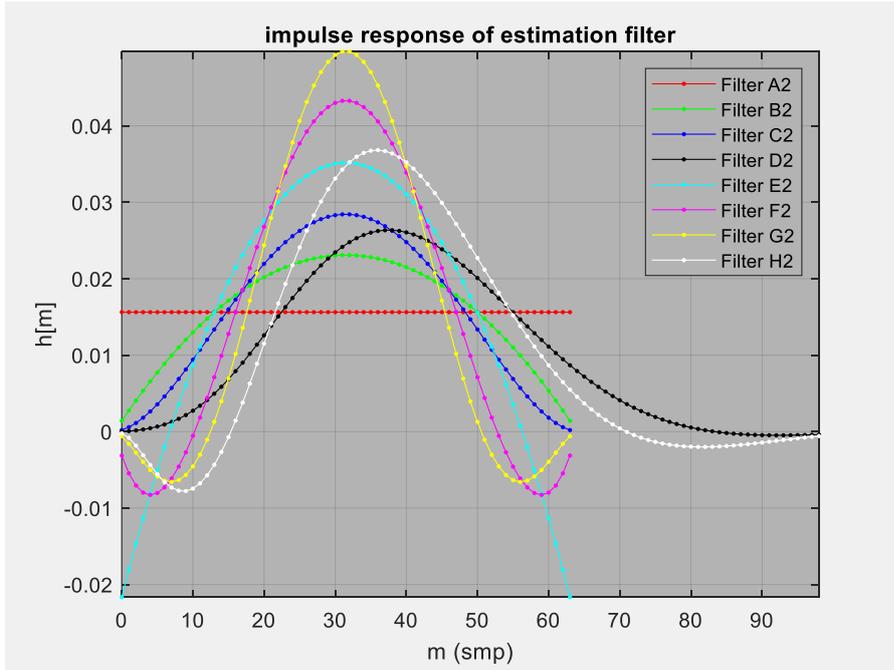

*Figure 8. Impulse responses of FIR & IIR estimators in Table 2.*

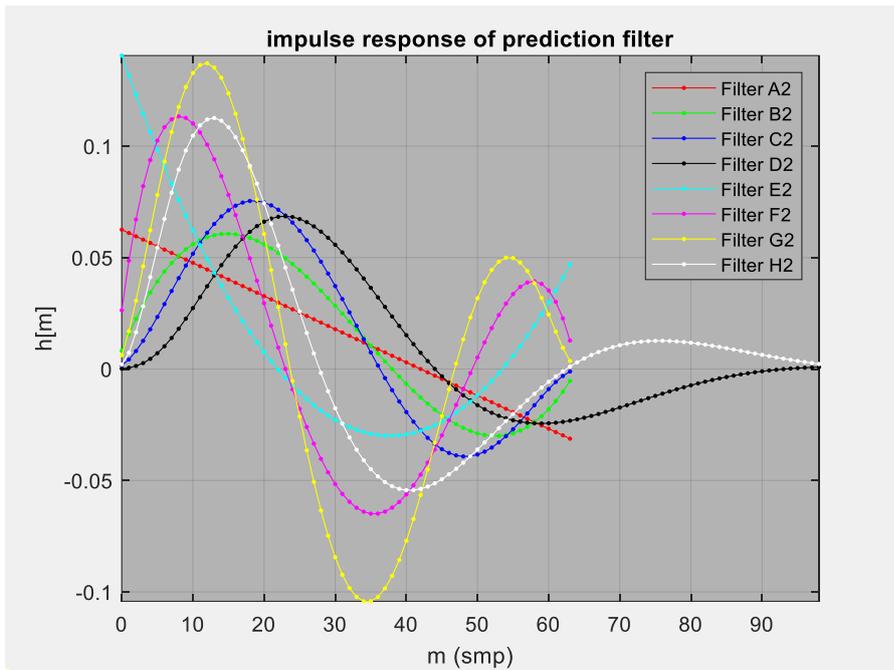

*Figure 9. Impulse responses of FIR & IIR predictors in Table 2.*

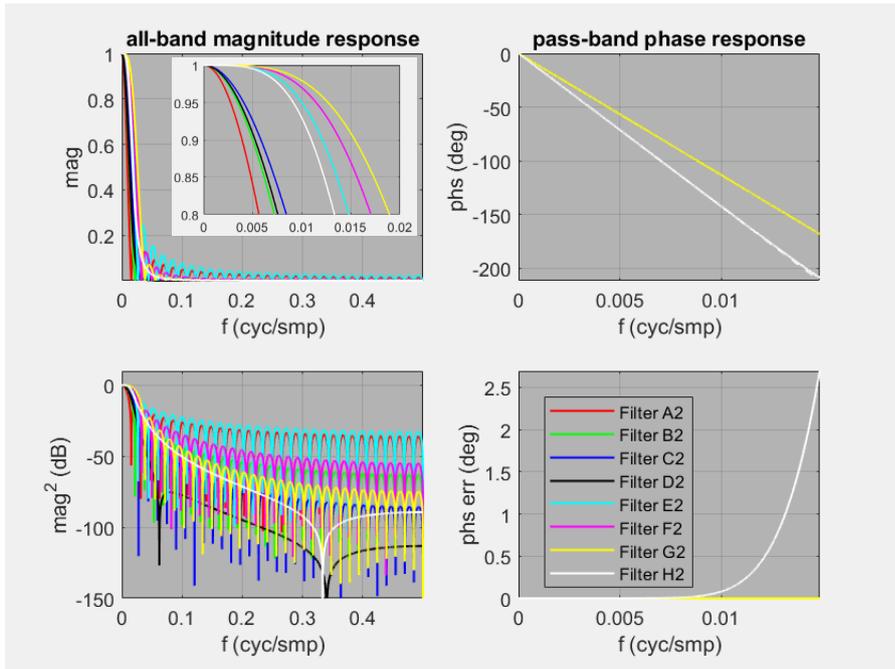

*Figure 10. Frequency responses of FIR & IIR estimators in Table 2.*

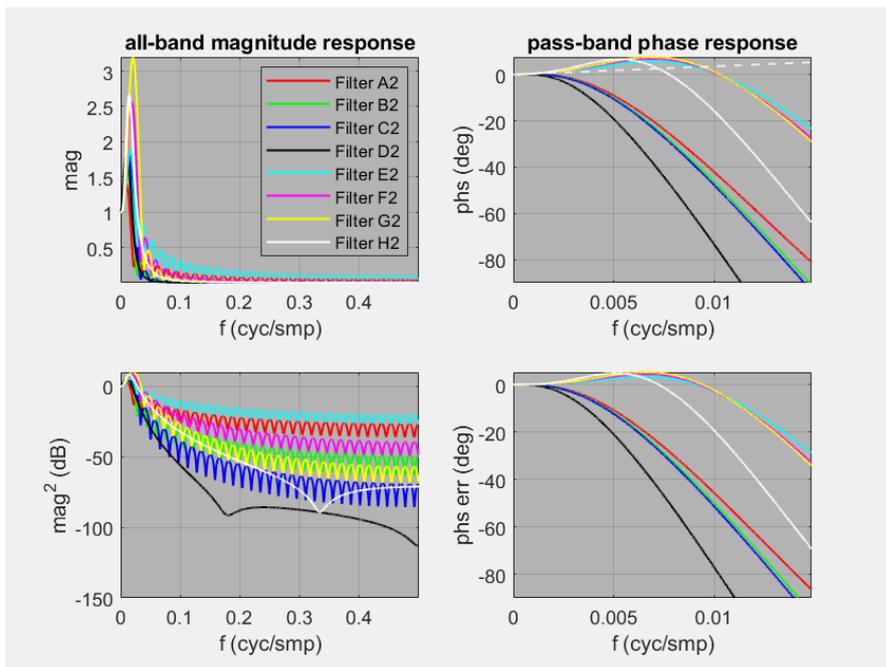

*Figure 11. Frequency responses of FIR & IIR predictors in Table 2.*

*Table 3. IIR filters used for signal type 2 with system configurations 1 & 2.*

| LPF ID | LPF Type | $K_1$ | $K_0$ | $M$ | $f_c$ | $q$ | $v_{LPF}$ | $v_{BPF}$ | $\tilde{v}_{BPF}$ ($\tilde{R}_0 = 1$) |
|--------|----------|-------|-------|-------|--------|--------|---------|-----------|----------------------------------------|
| A3 | IIR | 2 | 3 | 91.43 | 0.0109 | 56.800 | 0.0137 | 3.244E-10 | 1.520E-05 |
| B3 | IIR | 2 | 3 | 80.00 | 0.0125 | 49.643 | 0.0156 | 8.049E-10 | 2.248E-05 |
| C3 | IIR | 2 | 3 | 71.11 | 0.0141 | 44.078 | 0.0175 | 1.789E-09 | 3.170E-05 |
| D3 | IIR | 2 | 3 | 64.00 | 0.0156 | 39.626 | 0.0194 | 3.644E-09 | 4.306E-05 |
| E3 | IIR | 2 | 3 | 58.18 | 0.0172 | 35.984 | 0.0213 | 6.920E-09 | 5.676E-05 |
| F3 | IIR | 2 | 3 | 53.33 | 0.0188 | 32.950 | 0.0231 | 1.240E-08 | 7.298E-05 |
| G3 | IIR | 2 | 3 | 49.23 | 0.0203 | 30.383 | 0.0250 | 2.116E-08 | 9.189E-05 |

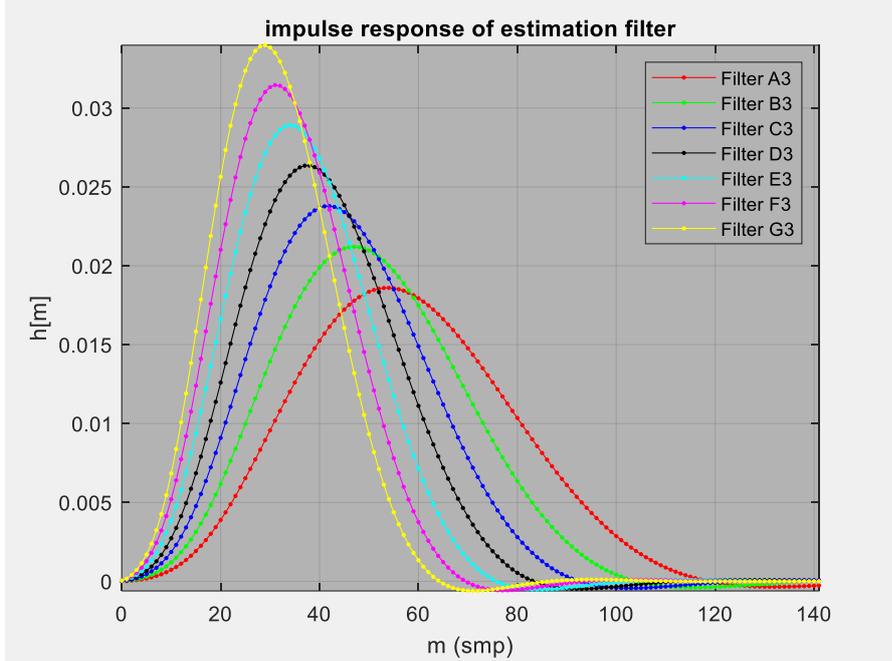

*Figure 12. Impulse responses of IIR estimators in Table 3.*

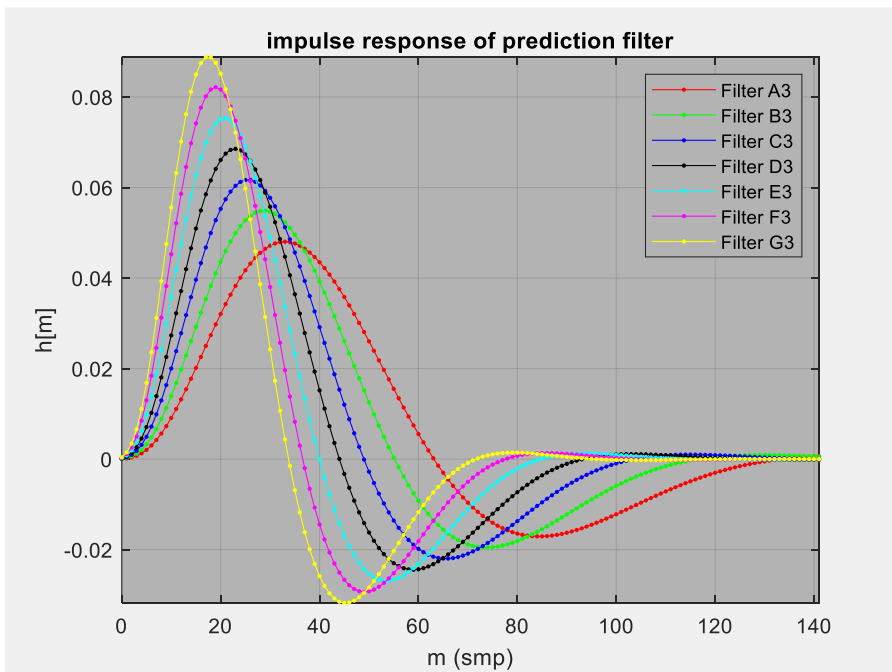

*Figure 13. Impulse responses of IIR predictors in Table 3.*

*Table 4. IIR filters used for signal type 2 with system configuration 3.*

| LPF ID | LPF Type | $K_1$ | $K_0$ | $M$ | $f_c$ | $q$ | $v_{LPF}$ | $v_{BPF}$ | $\tilde{v}_{BPF}$ ($\tilde{R}_0 = 0$) |
|--------|----------|-------|-------|-----|-------|-----|-----------|-----------|-------------|
| A4 | IIR | 3 | 3 | 128.00 | 0.0078 | 79.705 | 0.0156 | 3.416E-09 | 1.559E-02 |
| B4 | IIR | 3 | 3 | 106.67 | 0.0094 | 66.343 | 0.0187 | 1.017E-08 | 1.870E-02 |
| C4 | IIR | 3 | 3 | 91.43 | 0.0109 | 56.800 | 0.0218 | 2.556E-08 | 2.181E-02 |
| D4 | IIR | 3 | 3 | 80.00 | 0.0125 | 49.643 | 0.0249 | 5.676E-08 | 2.493E-02 |
| E4 | IIR | 3 | 3 | 71.11 | 0.0141 | 44.078 | 0.0280 | 1.146E-07 | 2.804E-02 |
| F4 | IIR | 3 | 3 | 64.00 | 0.0156 | 39.626 | 0.0311 | 2.149E-07 | 3.115E-02 |
| G4 | IIR | 3 | 3 | 58.18 | 0.0172 | 35.984 | 0.0343 | 3.790E-07 | 3.426E-02 |

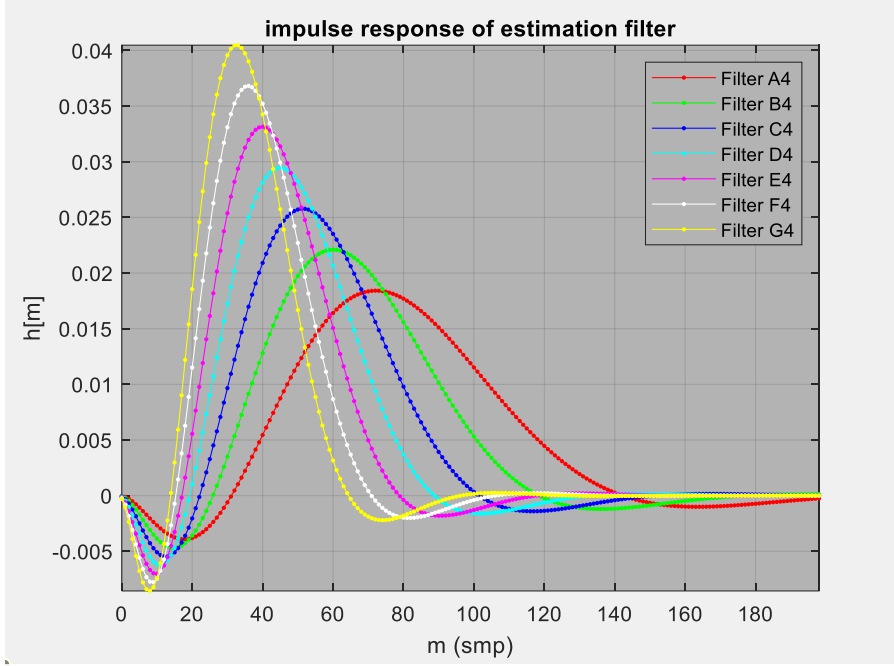

*Figure 14. Impulse responses of IIR estimators in Table 4.*

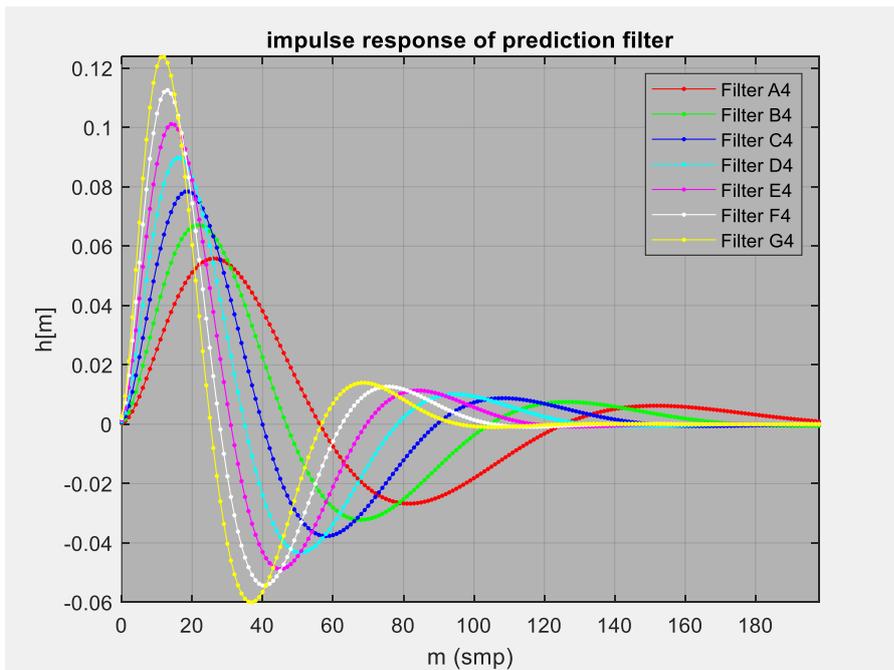

*Figure 15. Impulse responses of IIR predictors in Table 4.*

### 6.2.1 System configuration 1: Pre-differentiator

This configuration, with $\alpha = 1$ and $\beta = 0$, illustrates and verifies some of the theoretical results discussed in Section 3. The complex waveform output by the analogue-to-digital converter and the angle (i.e. frequency) measurement inputs to the low-pass filter are plotted in Figure 16. At high SNR, the results computed from the simulations results agree with the results expected from analysis (see Figure 17). The SNR threshold at which all filters begin to fail, due to incorrect unwrapping decisions, leading to simulation errors that are much greater than analysis errors, is around 10 dB. As expected from theory, all frequency estimators are unbiased at high SNR, because $K_1 \geq K_\theta - \widetilde{R}_0 + 1$, with $K_\theta = 2$ & $\widetilde{R}_0 = 1$ in this scenario. For the Filter A2-C2 series, all FIR with $K_1 = 2$ and $K_0 = 0$, 1 & 2, respectively, the estimator variance is minimised using $K_0 = \widetilde{R}_0 = 1$, i.e. for Filter B2, as expected from theory. Filter A2, with $K_0 = 0$ for a rectangular impulse response (see Figure 8), provides insufficient noise attenuation at high frequencies (see Figure 10). Filter C2, with $K_0 = 2$ for a more tapered impulse response, has much lower sidelobes but a wider mainlobe width (see Figure 10 inset), thus it provides insufficient noise attenuation at low frequencies. For the Filter E2-G2 series, all FIR with $K_1 = 3$ and $K_0 = 0$, 1 & 2, respectively, the main lobe is even wider than Filter C2, thus their variance is also elevated, relative to the optimal Filter B2. Filter D2, with $K_1 = 2$ & $K_0 = 3$ and an IIR with $f_c = 1/64$, has an error variance profile that is very similar to that of Filter B2. For the Filter A3-G3 series, all IIR with $K_1 = 2$ and $K_0 = 3$, the error variance at high SNR decreases with the filter bandwidth (Table 3) and the SNR threshold remains around 10 dB (see Figure 18).

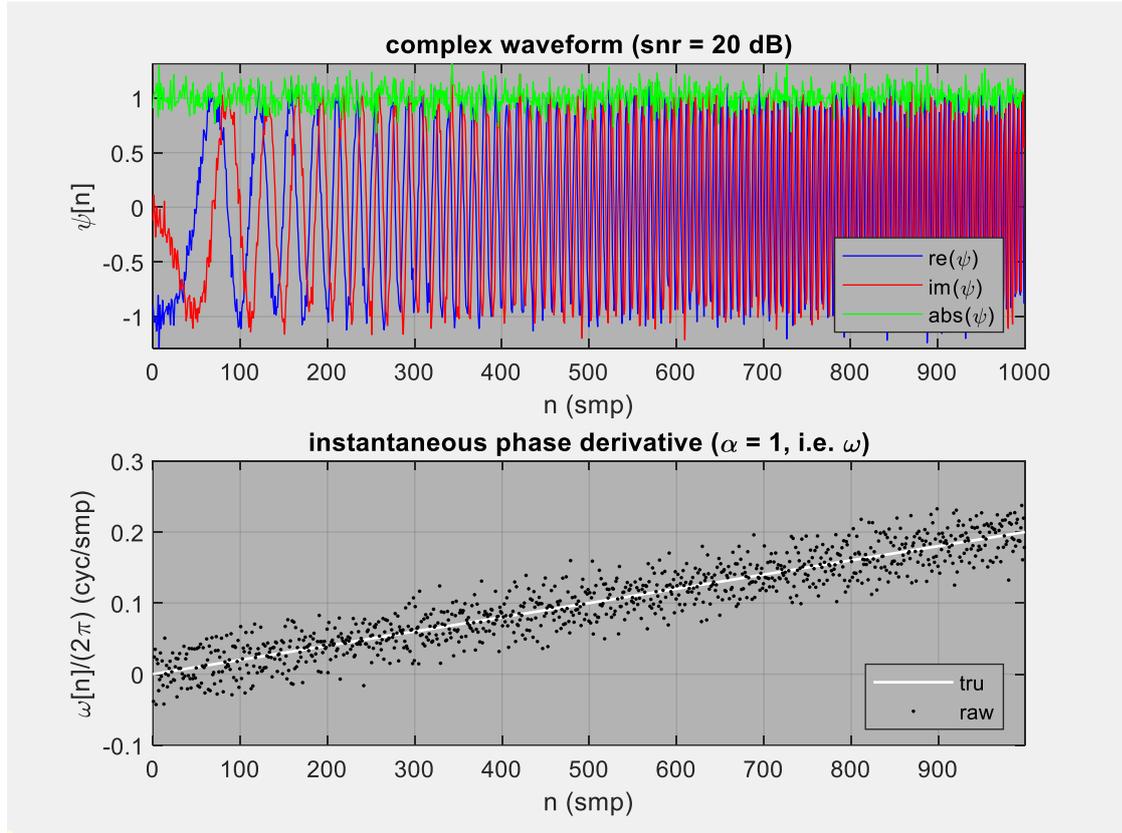

*Figure 16. Example ADC output and LPF input for signal type 2 and system configuration 1.*

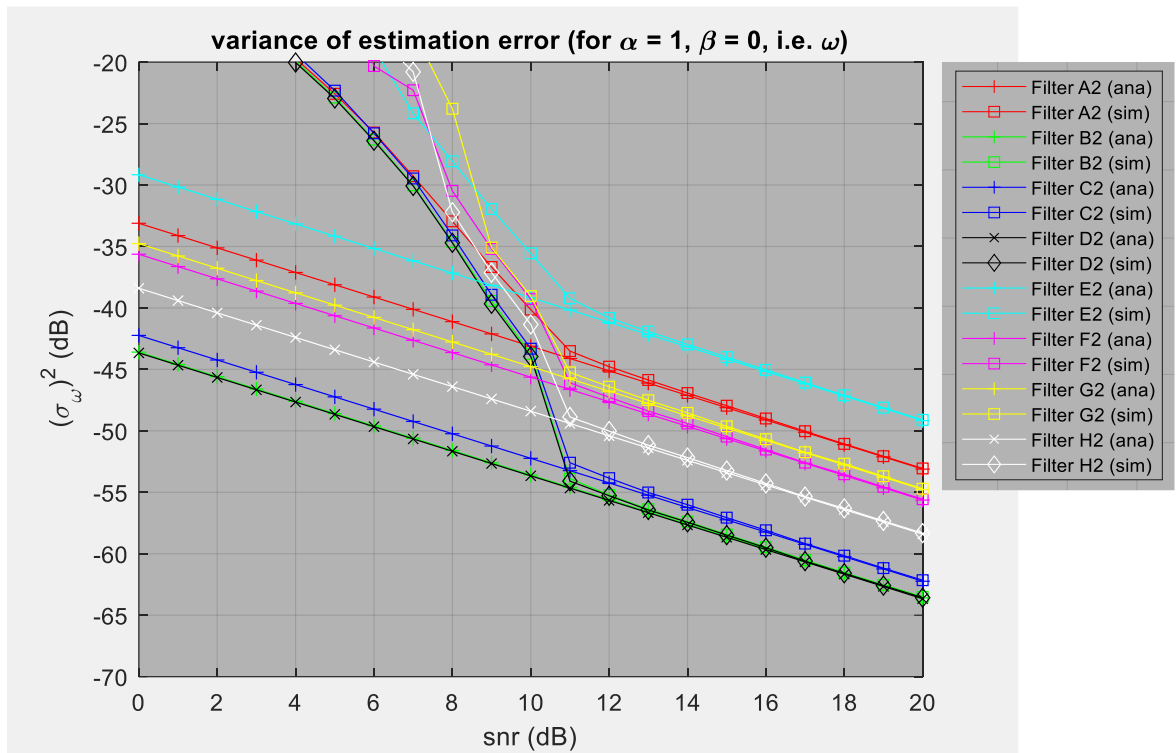

*Figure 17. Error variance versus SNR for signal type 2 and system configuration 1 using the FIR & IIR filters in Table 2*

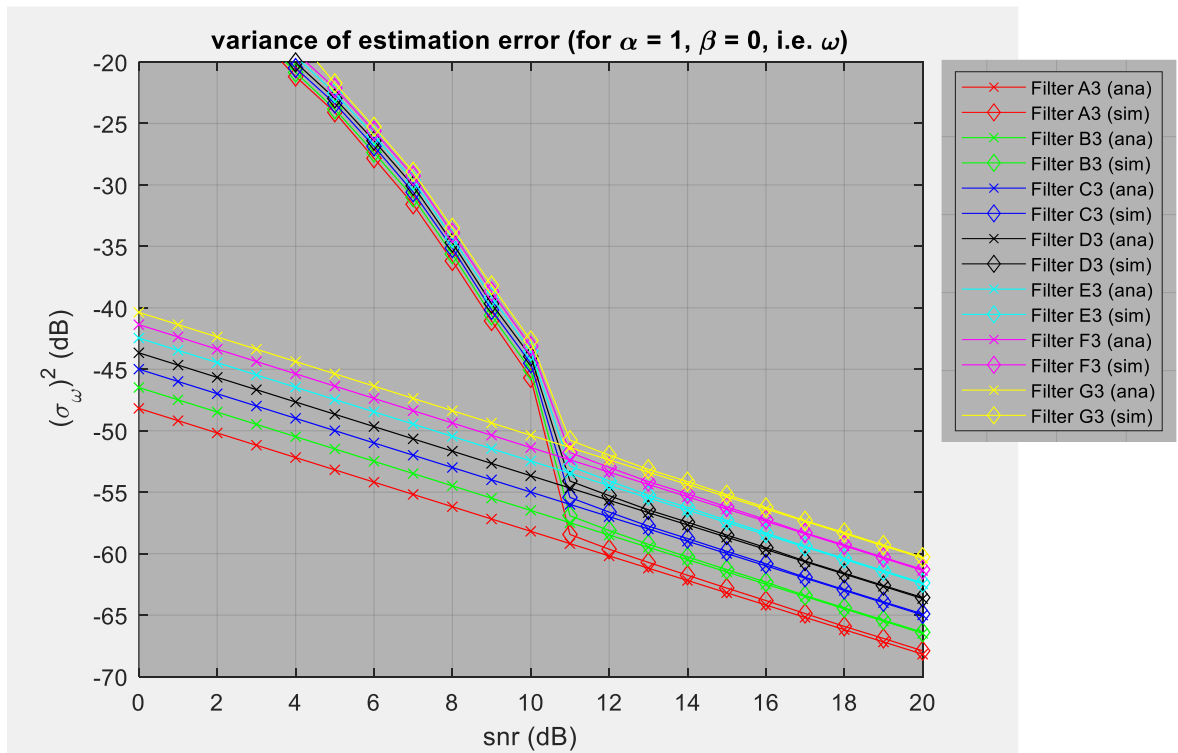

*Figure 18. Error variance versus SNR for signal type 2 and system configuration 1 using the IIR filters in Table 3.*

### 6.2.2 System configuration 2: Aft-differentiator

The configuration employed in this scenario applies the differentiator after the LPF (i.e. $\alpha = 0$ and $\beta = 1$) thus the LPF operates on raw phase measurements, instead of raw frequency measurements (see Figure 19). At high SNR, the results for configurations 1 & 2, using the filters in Table 2, are identical and the simulated results agree with the results expected from analysis (see Figure 17 & Figure 20); however, all filters now fail at a lower SNR (less than 10 dB). Filters A2 and E2 with the highest sidelobes (see Figure 10) and the highest CNG (see $\tilde{v}_{\mathrm{BPF}}$ in Table 2), thus the largest error variance at high SNR, have the lowest failure threshold (around 3 dB). The results for IIR Filters A3-G3 (see Figure 21) also indicate that filters with a wider bandwidth and a higher CNG (see Table 3), thus a higher error variance at high SNR, have a lower SNR failure threshold. It is likely that the degraded phase unwrapping performance for narrow-band filters is due to their longer impulse responses (see Figure 12 & Figure 13), which increases the time required for the LPF to converge after initialization or after a one-off phase unwrapping error. Lagged phase estimates that fall behind the accelerating measurement sequence are more likely to latch onto measurements that have wrapped around (see Figure 19).

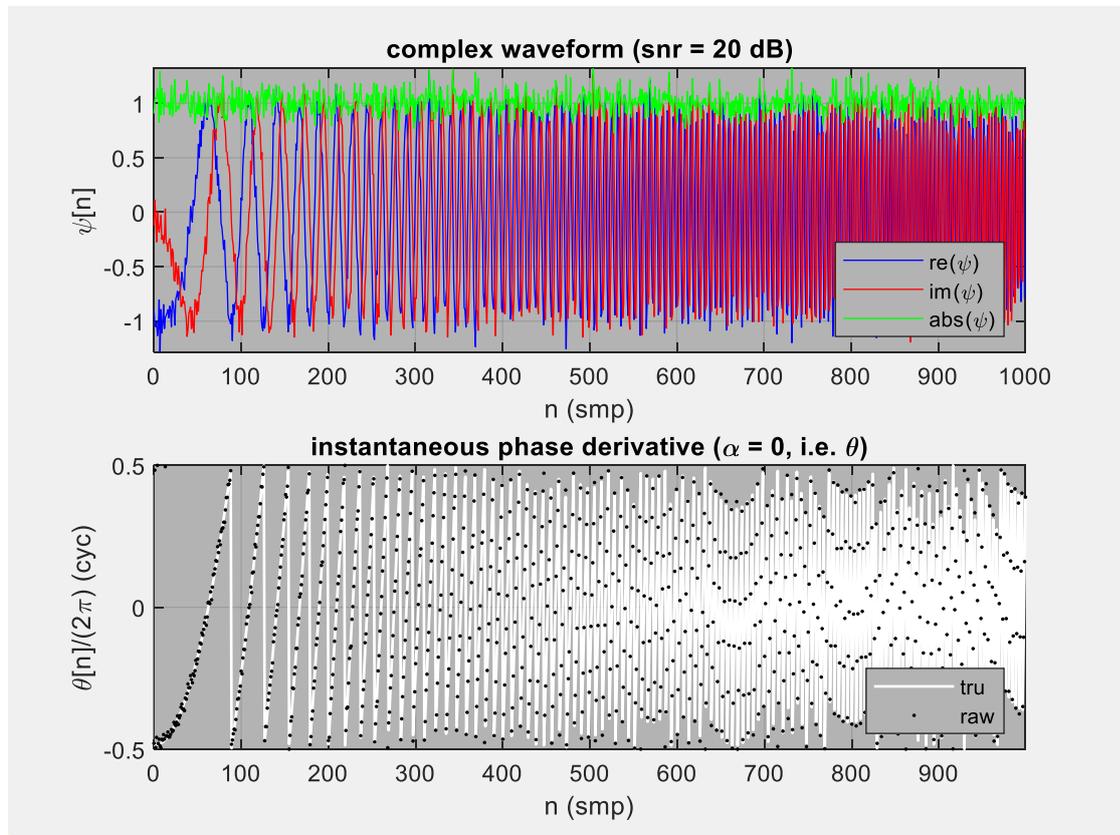

*Figure 19. Example ADC output and LPF input for signal type 2 and system configurations 2 & 3.*

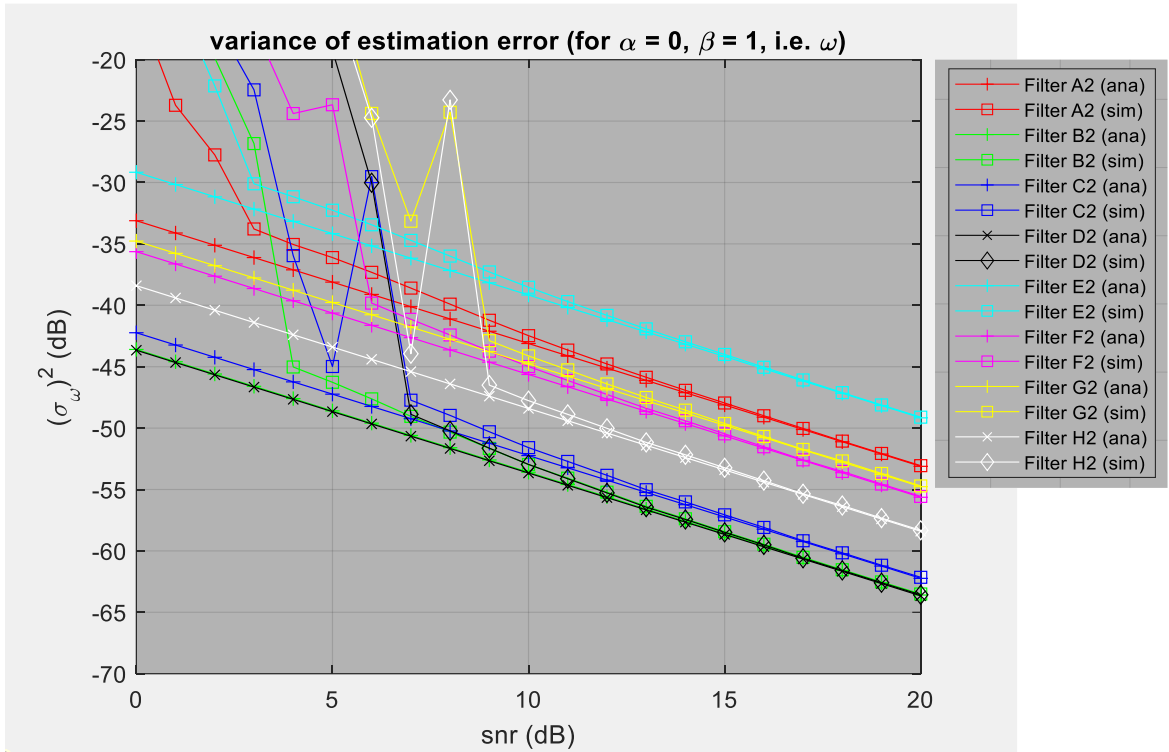

*Figure 20. Error variance versus SNR for signal type 2 and system configuration 2 using the FIR & IIR filters in Table 2.*

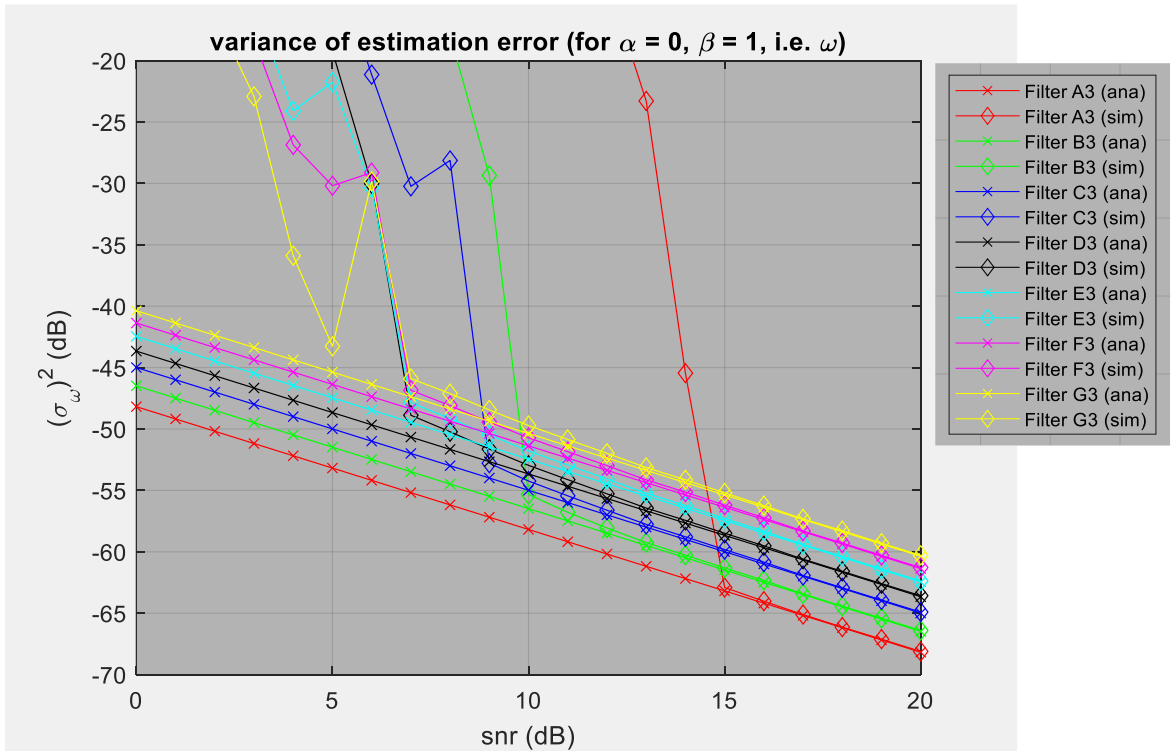

*Figure 21. Error variance versus SNR for signal type 2 and system configuration 2 using the IIR filters in Table 3.*

### 6.2.3 System configuration 3: Nil-differentiator

In this scenario, the LPF also operates on raw phase measurements (see Figure 19); however, the instantaneous phase is estimated instead of the instantaneous frequency, thus no differentiators are applied (i.e. $\alpha = 0$ and $\beta = 0$). The results show that Filters A2-D2 (all with $K_1 = 2$) are severely biased at high SNR because the condition $K_1 \geq K_\theta - \widetilde{K}_0 + 1$, is not satisfied in this scenario, with $K_\theta = 2$ & $\widetilde{K}_0 = 0$. For Filters E2-H2 (all with $K_1 = 3$) the simulated results are unbiased at high SNR and agree with the analysis results, and as expected from the theory discussed in Section 3, the error variance is lowest for Filter E2 with $K_0 = \widetilde{K}_0 = 0$ and greatest for Filter G2 with $K_0 = 2$. The unbiased filters all become biased below 10 dB, where phase unwrapping errors dominate. Filters B4-G4 in Table 4 (all IIR with $K_1 = 3$ & $K_0 = 3$) are also all unbiased at high SNR, and as observed for configuration 2, the error variance at high SNR decreases with the filter bandwidth; however, for this configuration, the SNR failure threshold is roughly the same for all filters (see Figure 23). At high SNR, the simulation error variance of Filter A4 (with the narrowest bandwidth thus the longest timescale) is greater than the analysis error variance. When the simulation error variance is computed over $n = (N-1)/4 \ldots (N-1)$ instead of $n = (N-1)/8 \ldots (N-1)$ the error variance gap is removed, which suggests that the elevated error variance is due to the longer startup transient, and the associated bias, for this narrowband filter.

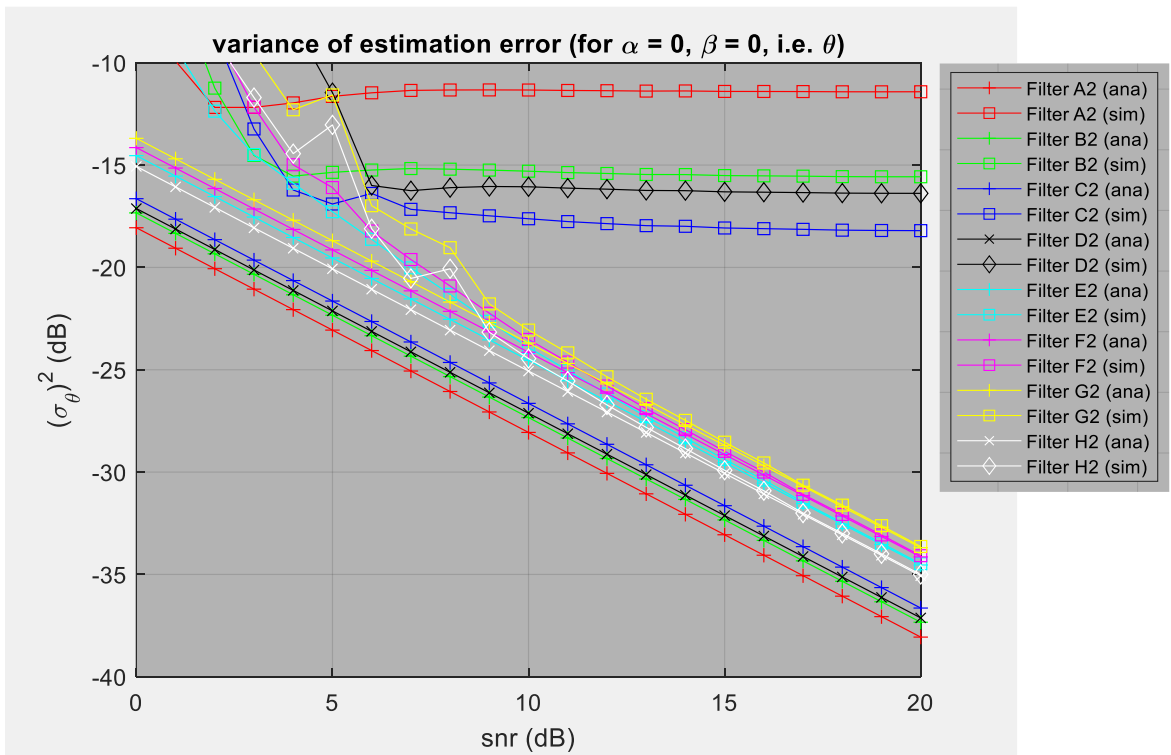

Figure 22. Error variance versus SNR for signal type 2 and system configuration 3 using the FIR & IIR filters in Table 2.

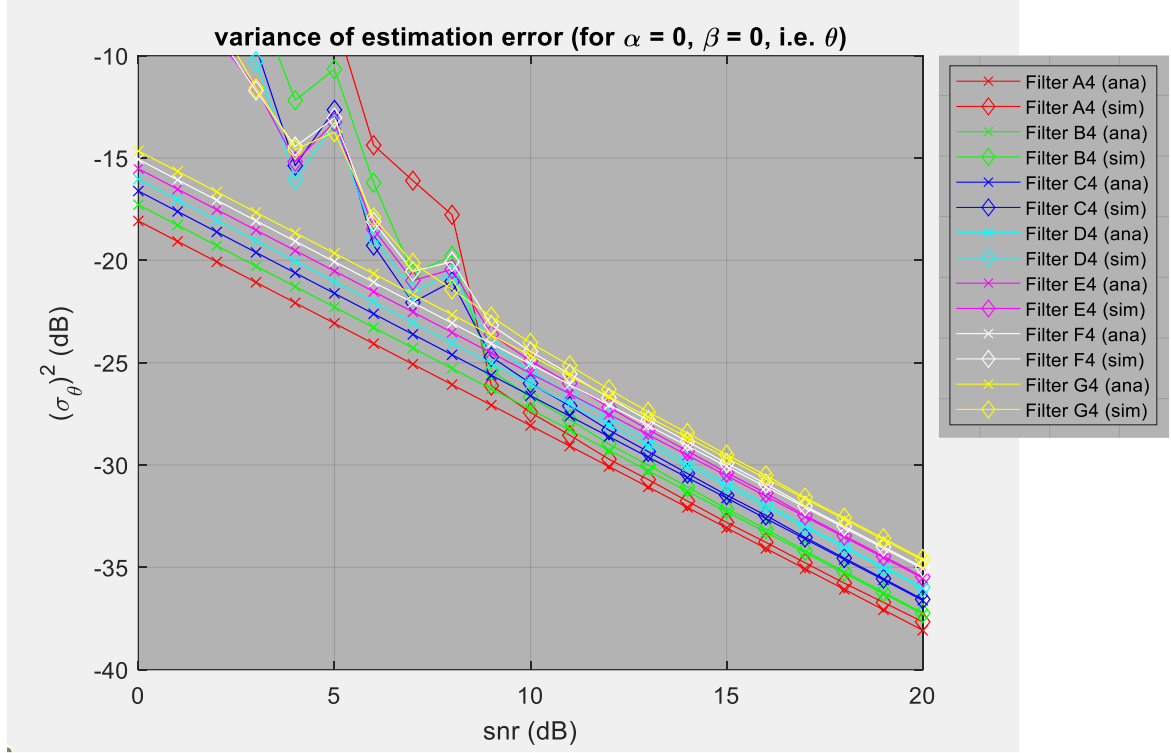

*Figure 23. Error variance versus SNR for signal type 2 and system configuration 3 using the IIR filters in Table 4.*

## 6.3  Signal type 3: Quadratic frequency sweep

A signal with a (wideband) quadratic frequency sweep (i.e. $K_\theta = 3$) with $f_0 = 0.0$, $f_1 = 4.0$ & $f_2 = 16.0$ is considered here. The parameters of the low-pass filters used for this scenario are provided in Table 5. Filter A5 is a simple rectangular window (with $K_1 = 1$ & $K_0 = 0$); all other filters were designed using $K_1 = 3$. Filters B5, C5 & D5 are FIR with $K_0 = 1, 2$ & 3. Filters E5, F5 & G5 were designed to be their IIR 'equivalents' (also with $K_0 = 1, 2$ & 3). The CNG ($v_{BPF}$) of each IIR filter was matched to its corresponding FIR filter. This was manually done by adjusting the bandwidth of the IIR basis set (i.e. $f_c$) until the CNG of the FIR and IIR estimators matched to four significant figures. The responses of the resulting (FIR and IIR) estimators and predictors are plotted in Figure 24, Figure 25, Figure 26 & Figure 27.

*Table 5. FIR and IIR filters used for signal type 3.*

| LPF ID | LPF Type | $K_1$ | $K_0$ | $M$ | $f_c$ | $q$ | $v_{LPF}$ | $v_{BPF}$ | $\tilde{v}_{BPF}$ ($\tilde{K}_0 = 1$) | $\tilde{v}_{BPF}$ ($\tilde{K}_0 = 0$) |
|---|---|---|---|---|---|---|---|---|---|---|
| A5 | FIR | 1 | 0 | 64.00 | 0.0156 | 31.500 | 0.0156 | 1.563E-02 | 4.883E-04 | 1.563E-02 |
| B5 | FIR | 3 | 1 | 64.00 | 0.0156 | 31.500 | 0.0385 | 2.734E-04 | 2.734E-04 | 3.851E-02 |
| C5 | FIR | 3 | 2 | 64.00 | 0.0156 | 31.500 | 0.0427 | 7.061E-06 | 3.334E-04 | 4.272E-02 |
| D5 | FIR | 3 | 3 | 64.00 | 0.0156 | 31.500 | 0.0465 | 3.385E-07 | 4.272E-04 | 4.653E-02 |
| E5 | IIR | 3 | 1 | 51.02 | 0.0196 | 31.500 | 0.0385 | 2.734E-04 | 2.734E-04 | 3.851E-02 |
| F5 | IIR | 3 | 2 | 51.02 | 0.0196 | 31.501 | 0.0390 | 7.061E-06 | 2.828E-04 | 3.902E-02 |
| G5 | IIR | 3 | 3 | 59.30 | 0.0169 | 36.682 | 0.0336 | 3.385E-07 | 1.815E-04 | 3.361E-02 |

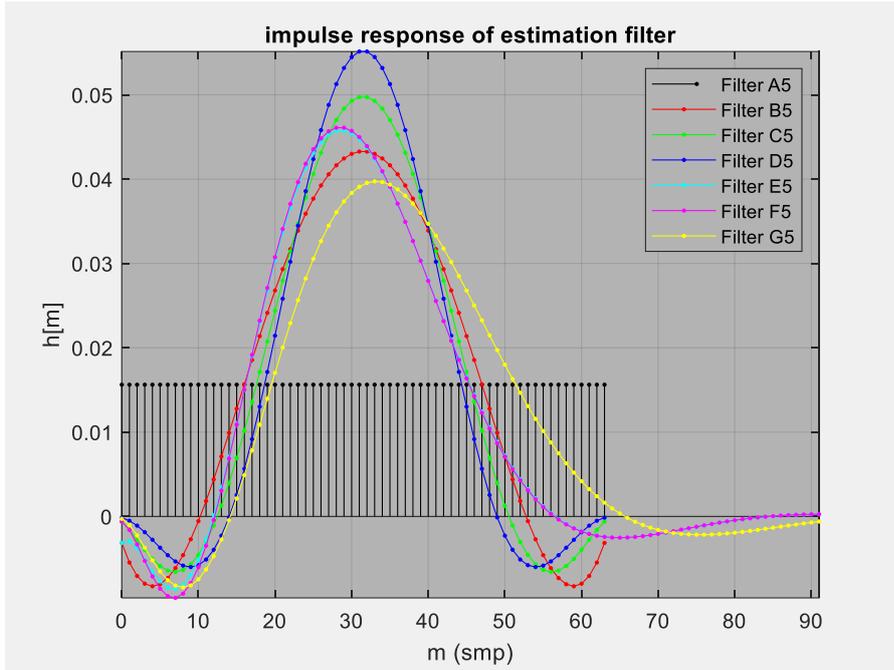

*Figure 24. Impulse responses of FIR & IIR estimators in Table 5.*

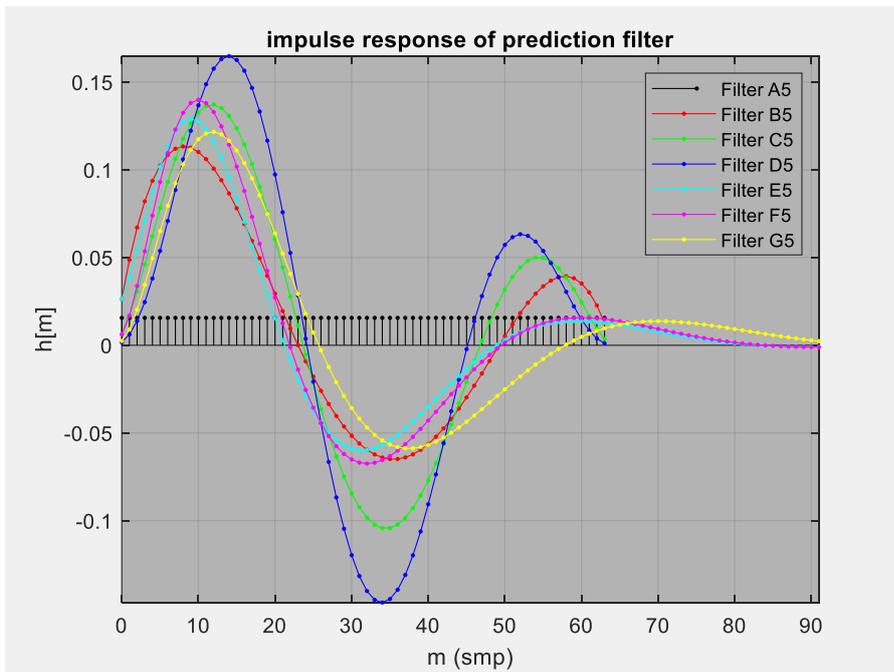

*Figure 25. Impulse responses of FIR & IIR predictors in Table 5.*

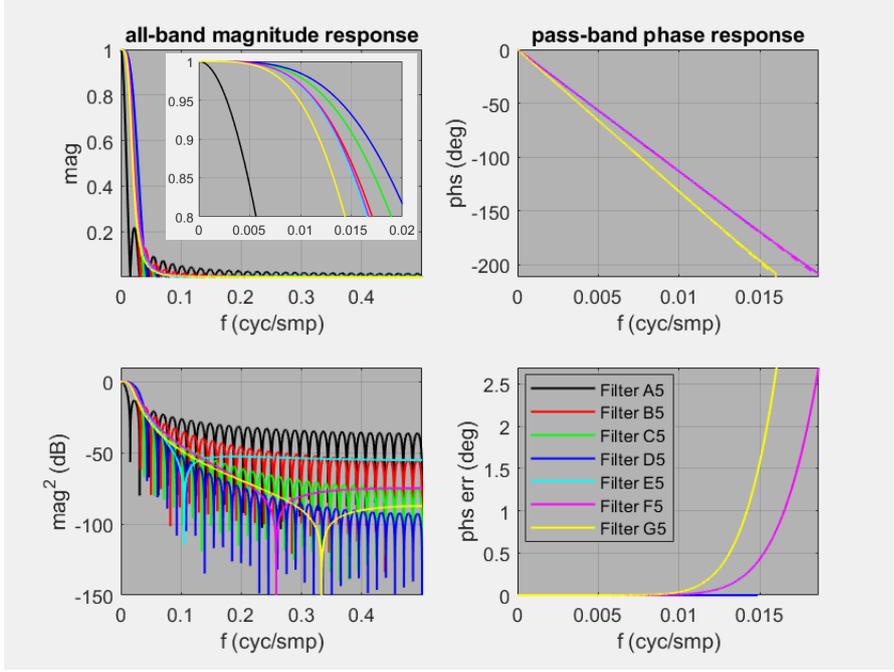

*Figure 26. Frequency responses of FIR & IIR estimators in Table 5.*

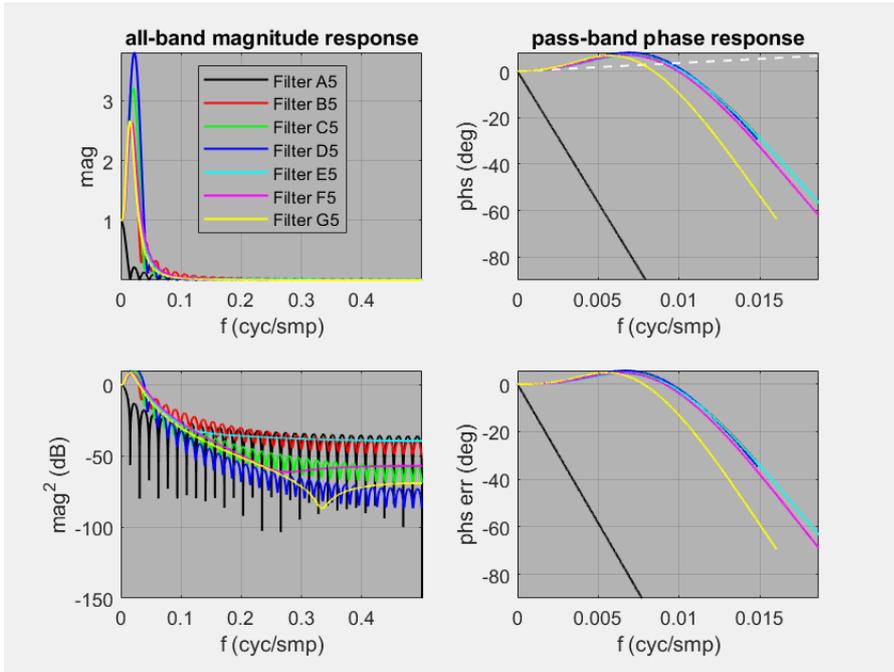

*Figure 27. Frequency responses of FIR & IIR predictors in Table 5.*

### 6.3.1 System configuration 1: Pre-differentiator

The FIR and IIR filters designed using parameters that are matched to the simulation (i.e. Filters B5 & E5, with $K_1 = 3$ & $K_0 = \widetilde{K}_0 = 1$) yield identical (simulated and analysis) results at high SNR. For the FIR and IIR filters designed with mis-matched parameters, the bandwidth of the IIR filter is less than the corresponding FIR filter (see Figure 26) which yields a lower error variance at high SNR (see $\tilde{v}_{\text{BPF}}$ in Table 5). At low SNR, Filters B5-G5 (all with $K_1 = 3$) fail at around the same SNR threshold (around 10 dB). Filter A5 with $K_1 = 1$ has an error variance of around -6 dB for all SNR (not within the range of the vertical axis in Figure 26).

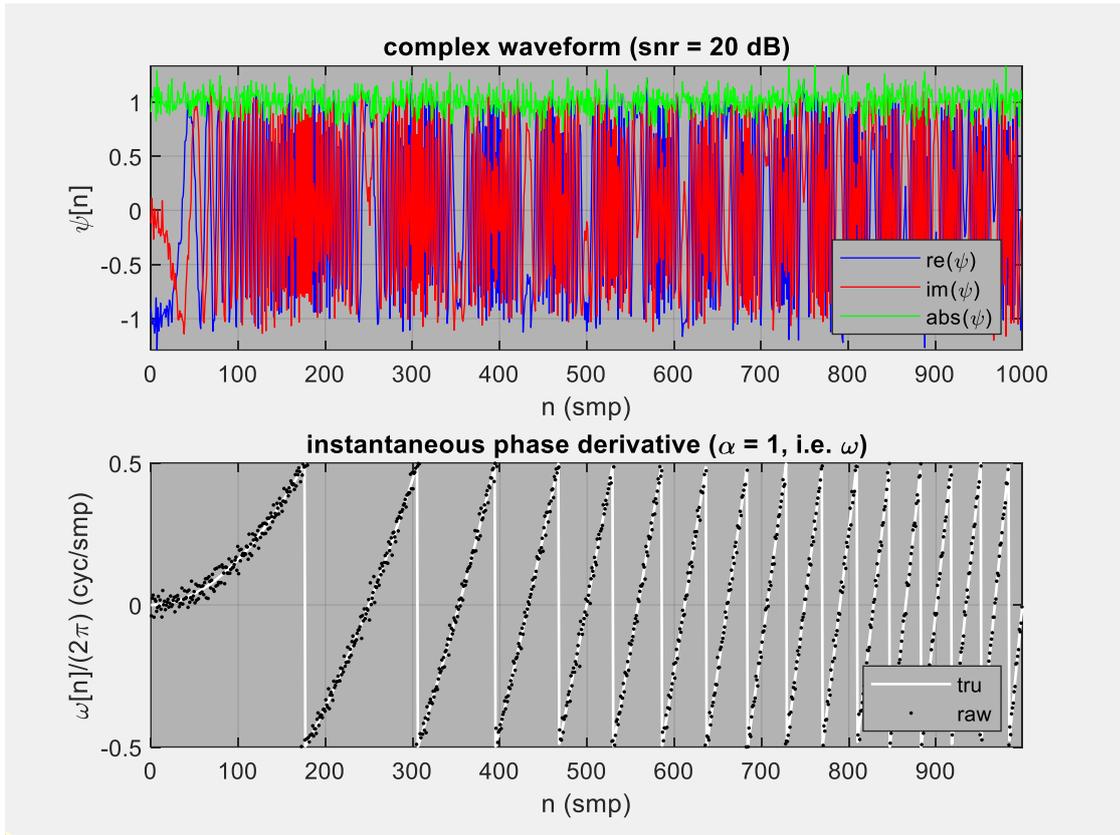

*Figure 28. Example ADC output and LPF input for signal type 3 and system configuration 1.*

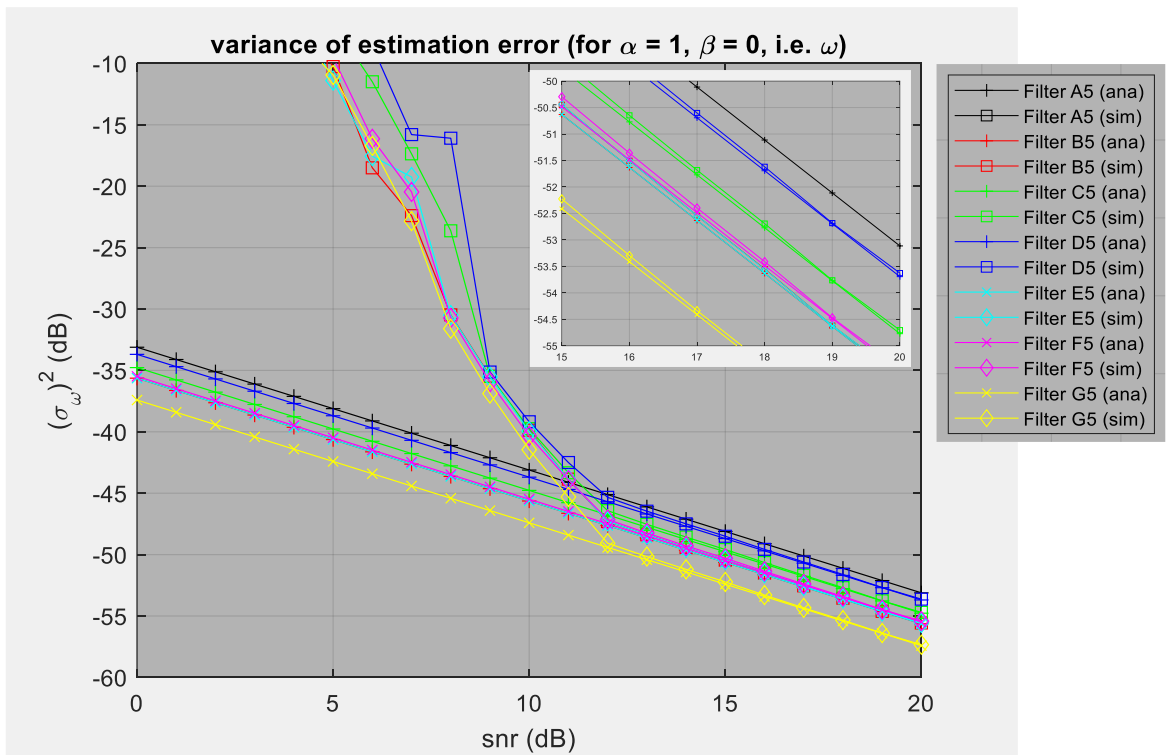

*Figure 29. Error variance versus SNR for signal type 3 and system configuration 1 using the FIR & IIR filters in Table 5.*

### 6.3.2 System configuration 2: Aft-differentiator

As was the case for signal type 2, in this scenario, the predictors with a phase lead (for phase unwrapping) allow the estimators with a phase lag (for low error variance) to operate down to a lower SNR (see Figure 31). FIR Filter B5, and IIR filters E5 & F5, all fail around 6 dB. The analysis error variance of each filter does not change when the differentiator is move from pre (configuration 1) to aft (configuration 2); the simulated error variance is the approximately the same at high SNR.

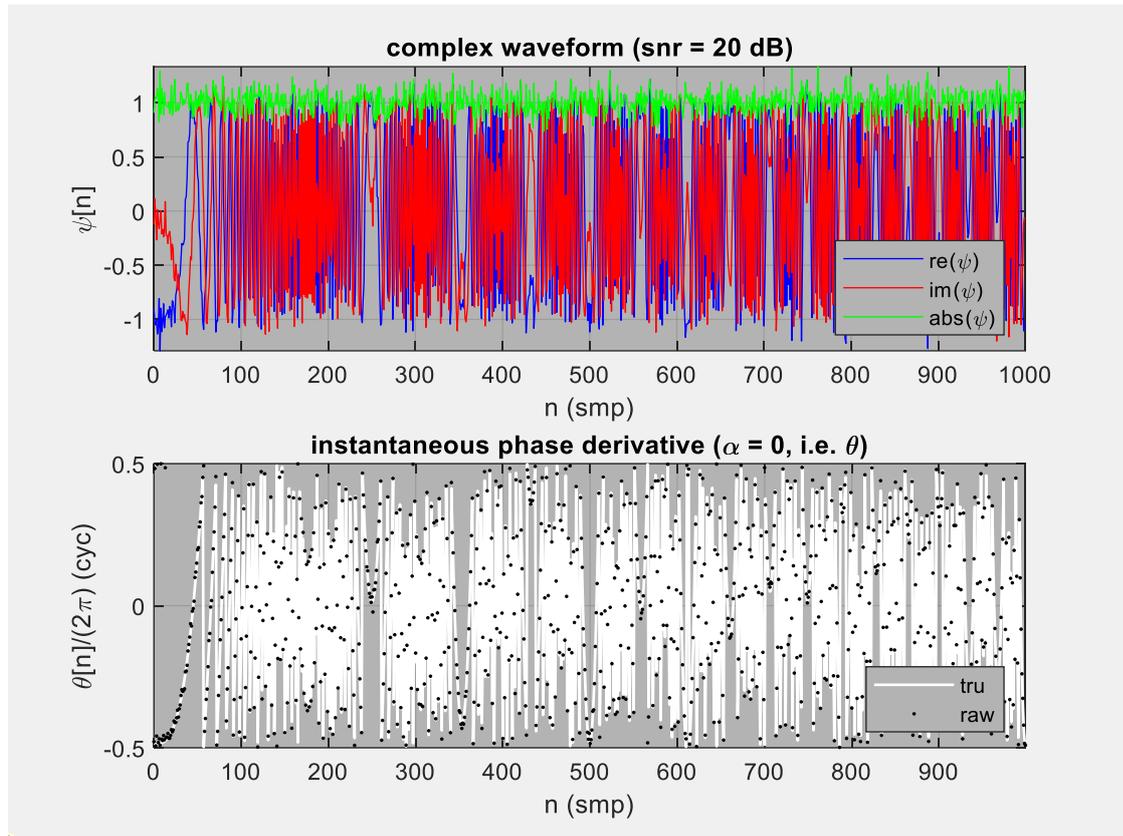

*Figure 30. Example ADC output and LPF input for signal type 3 and system configurations 2 & 3.*

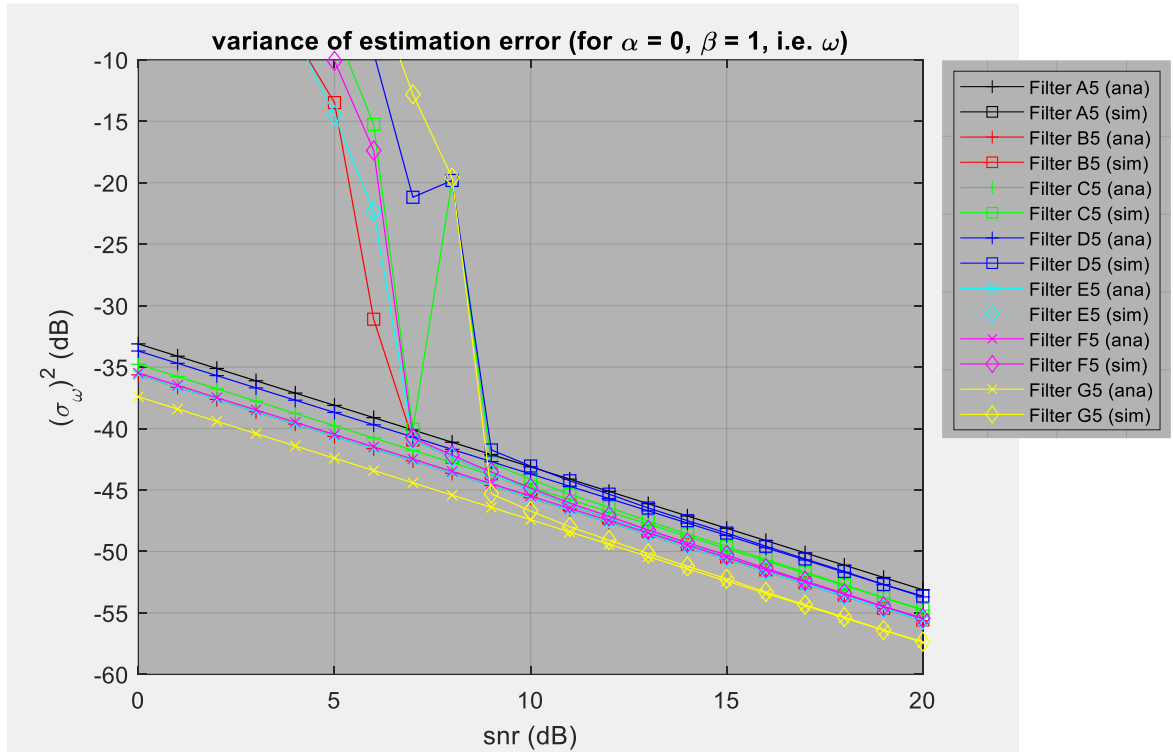

*Figure 31. Error variance versus SNR for signal type 3 and system configuration 2 using the FIR & IIR filters in Table 5.*

### 6.3.3   System configuration 3: Nil-differentiator

No differentiators are applied in this configuration (i.e. $\alpha = 0$ and $\beta = 0$) because estimates of instantaneous phase are required. In theory, the estimates produced by Filters B5-G5 in Table 5 should be biased because the condition $K_1 \geq K_\theta - \widetilde{K}_0 + 1$, is not satisfied for this scenario, with $K_\theta = 3$ & $\widetilde{K}_0 = 0$. However, the simulated results (see Figure 32) suggest that all filters are unbiased at high SNR. The phase polynomial of this signal (with a quadratically swept frequency) has a very small cubic coefficient; thus, the bias is very small and only visible at very high SNR when the random errors are also very small (see Figure 33).

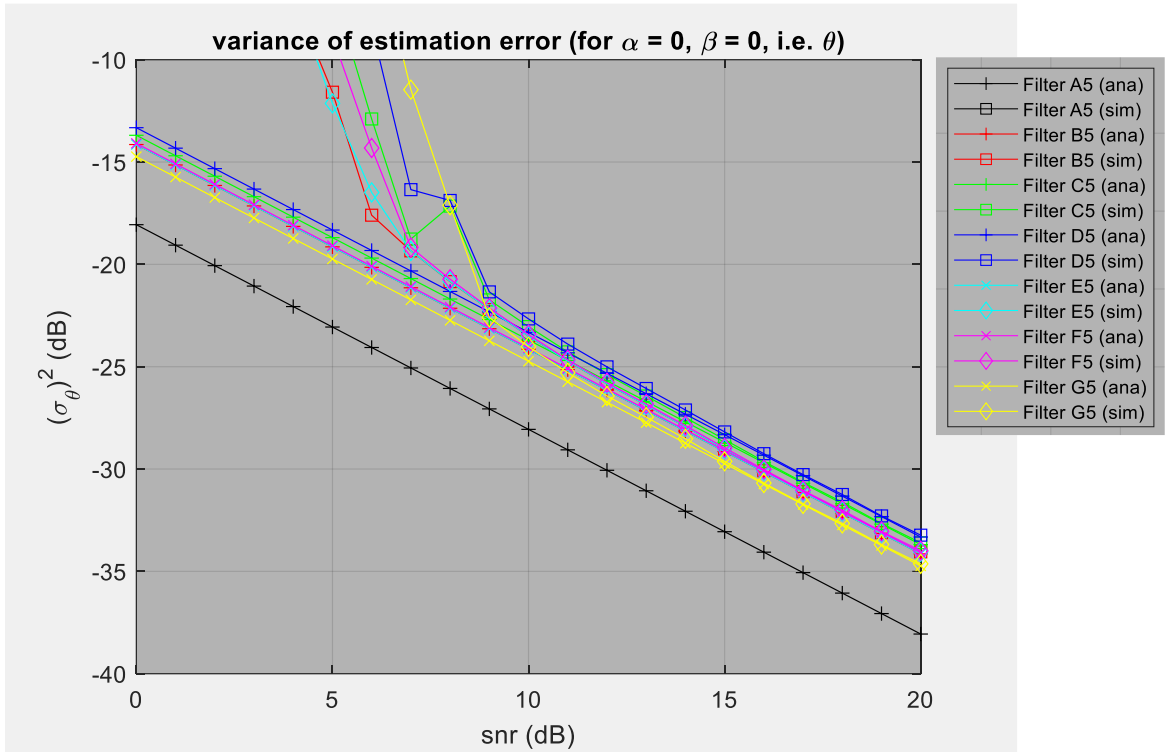

*Figure 32. Error variance versus SNR for signal type 3 and system configuration 3 using the FIR & IIR filters in Table 5.*

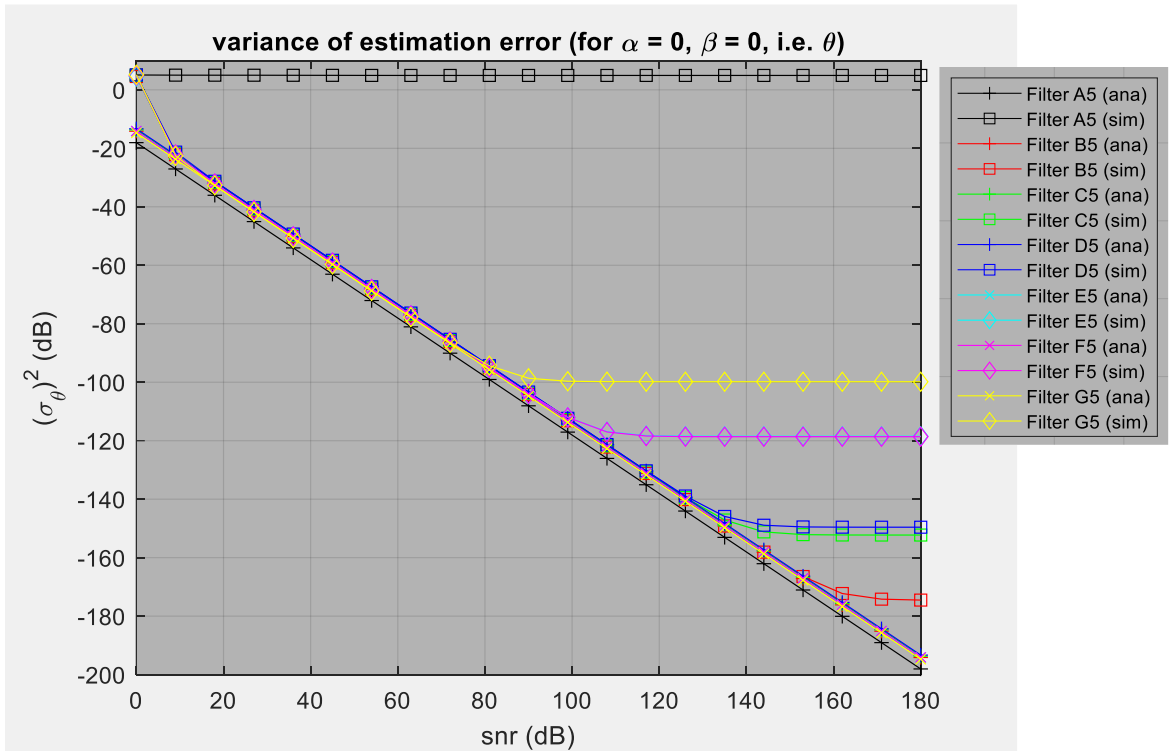

*Figure 33. Error variance versus SNR for signal type 3 and system configuration 3 using the FIR & IIR filters in Table 5, at very high SNR.*

# 7   Discussion

It was stated in Section 3 that the additive measurement noise ($\varepsilon_\psi$) is assumed only to be white; however Gaussian distributed errors were used in the MC simulations of Section 6. To test this statement, the simulations were repeated using (zero-mean) uniformly distributed measurement errors, i.e. $\mathrm{Re}\{\varepsilon_\psi\} \sim \mathcal{U}(0, \sigma_\varepsilon^2)$ and $\mathrm{Im}\{\varepsilon_\psi\} \sim \mathcal{U}(0, \sigma_\varepsilon^2)$. Visually identical results were observed at high SNR (e.g. see Figure 34 & Figure 35), i.e. the simulated results agreed with analysis results; however, slightly lower failure thresholds were typically observed, presumably due to the longer tails of the Gaussian distribution, which are more likely to cause angle unwrapping errors.

It was also asserted in Section 3 that for white complex errors $\varepsilon_\psi$, the angle errors $\varepsilon_\theta$, are approximately white, with $\varepsilon_\theta[n] \approx \varepsilon_\psi[n]/A$, and that at low SNR "unmodelled correlations/distortions increase the likelihood of phase unwrapping errors". To test this assertion, the simulations in Section 6.2.1 were repeated with randomly generated (real) errors $\varepsilon_\theta[n]$, added directly to $\theta[n]$, with $\varepsilon_\theta[n] \sim \mathcal{N}(0, \sigma_\varepsilon^2/A^2)$, rather than adding randomly generated (complex) errors $\varepsilon_\psi[n]$ to $\psi[n]$. At high SNR, there is no change to the error variance of the estimator. At low SNR, this modification lowers the failure threshold for configuration 1, from around 10 dB to around 6 dB (see Figure 36), which is similar to the results obtained for configuration 2. This suggests that the non-linear $\arg\{\blacksquare\}$ operator does indeed contribute to phase unwrapping errors at low SNR, particularly for differentiated inputs with amplified high-frequency noise.

Removing the differentiator that precedes the low-pass filter allows instantaneous phase to be estimated, e.g. for phase-shift measurement, or (ambiguous) time-delay measurement, using system configuration 3. Moving differentiator after the low-pass filter (i.e. system configuration 2) allows instantaneous frequency to be estimated, e.g. for Doppler-shift measurement, and lowers the failure threshold (relative to configuration 1) from around 10 dB to around 3-6 dB for the signals considered here.

Improving the performance of the frequency estimator down to a lower SNR, using an aft differentiator (configuration 2) instead of a pre differentiator (configuration 1), with a prediction filter for phase unwrapping, is one of the main results of this paper. In this modified configuration, unwrapping is done for an angle input with reduced noise amplification and distortion, which makes noise-induced unwrapping errors less likely; however, the low-pass filter must now track an input with a greater rate of change, which makes signal-induced unwrapping errors more likely. Smoothing filters with a long lag are good for reducing the error variance of the frequency estimates but they are bad for making angle unwrapping decisions, particularly for inputs with high rates of change; thus, a predictor with a one sample lead is required to realise the benefits of the aft configuration. The results in Section 6.1.2, suggest that this is the case; however, for further verification, the simulations in Section 6.2.2 were repeated, with the output of the estimator used for phase unwrapping instead of the output of the predictor. All filters failed to track the rapidly changing phase polynomial (see Figure 19) and the estimator variance was around -3 dB at all SNR (0 to 20 dB). This suggests that the proposed estimator/predictor pairing is indeed responsible for the lower SNR threshold observed for simulation scenarios involving configuration 2.

The simulation results in Section 6 (e.g. see Section 6.2.2) also suggest that the failure threshold may be further lowered for configuration 2 (with the aft differentiator) by increasing the bandwidth of the low-pass filters to shorten the duration of their transient responses. The simulations in Section 6.2 were therefore repeated using low-pass filters designed using $M = 32$ instead of $M = 64$, for a halved timescale thus a doubled bandwidth. The failure threshold for configuration 1 (see Figure 37) remains approximately the same (around 10 dB) for all filters because unwrapping errors are caused by noise. However, the failure threshold for configuration 2 (see Figure 38 & Figure 39) is lowered for all filters (down to around 3 dB in some cases) because these filters with a wider bandwidth, thus a shorter timescale, converge faster and are able to track inputs with higher rates of change with reduced bias. If a large bias develops during track establishment, the angle track (i.e. the estimate) is more likely to be 'seduced' by angle measurements that have wrapped around. For reasonable angle (rate) estimates down to a low SNR, the duration of the impulse response of the LPF (i.e. its timescale) should be less than the expected time for the angle input to wrap around, e.g. for phase measurement inputs, the inequality $f_c > f_\psi$ should be satisfied. For both configurations 1 & 2, the error variance of the estimators is increased at high SNR, because these filters with a wider bandwidth have a higher CNG; however, the threshold at which phase unwrapping fails is significantly lower for all filters. Thus, there is clearly a bias-error versus random-error trade-off to consider when designing a filter and configuring a system for phase and/or frequency estimation. The bandwidth should be tuned for the desired transient response (for fewer unwrapping errors at low SNR) and the desired steady-state response (for lower error variance at high SNR). A non-negligible bandwidth thus a non-infinite timescale are also necessary in practice, to accommodate sudden changes in the polynomial parameters.

There is no universal definition of bandwidth, although the -3 dB frequency is commonly used [37]. The frequency of the first null, is a convenient way of specifying the mainlobe or passband width of a rectangular smoothing window, which is at $f = 1/M$ for this $K_1 = 1$ & $K_0 = 0$ filter. From this primitive starting point, sidelobes are lowered, and the mainlobe is broadened using $K_0 > 0$. There is therefore a mainlobe/sidelobe trade-off. However, the mainlobe width and sidelobe height may be decreased by increasing the filter order (which is equal to $M - 1$). IIR filters are not constrained in the same way, because the mainlobe width and high-frequency gain, may both be reduced by decreasing the bandwidth parameter ($f_c$). Increasing the filter order only improves the response up to a point and there is not much to be gained from using high-order filters as matrix ill conditioning quickly becomes a problem for large $K_\varphi$. For both FIR and IIR filters, the passband flatness (thus width) increases with $K_1$.

The signals considered in Section 6.2 & Section 6.3 are relatively easy to process using the proposed low-pass filters in system configuration 2 (with an aft differentiator) because the frequency sweeps all start from zero. For both linear and quadratic sweeps, the initial rate of phase change is zero, but it increases rapidly with time. For the quadratic sweep, the rate of phase change stays lower for longer, but then becomes faster than the linear sweep after some time has elapsed. The problem is much more difficult if the initial frequency is non-zero, e.g. for down sweeps instead of up sweeps, because the low-pass filters must track a very high rate of angle change from the moment they are initialized, which is only possible for filters with a very narrow bandwidth. This may be a problem for passive sensing systems that intercept transmissions from non-cooperative emitters [35] & [36]. However, in active systems that receive and process their own transmissions (e.g. radars and ultrasounds) waveforms that are most amenable to processing may be arbitrary selected and emitted [37].

The literature on optimal low-pass filters for the estimation of instantaneous phase derivatives often refers to two classes of streaming estimators [1], [3] & [34]. In the first class, the LPF is applied in the (real) angle domain *after* the arg{■} operator has been applied. In the second class, the LPF is applied in the complex domain, to the sampled waveform, *before* the arg{■} operator is applied, i.e. to the delayed conjugate products $x[n]x^*[n-1]$ in (3.11b). The latter approach is adopted in [35] & [36], with a basic rectangular window (with a uniform weight) performing the low-pass filtering function. This paper deals exclusively with the former class of estimators, mainly because they reach the Cramer-Rao lower bound (CRLB) at high SNR and stay close to this bound as the SNR threshold is approached. The latter class of estimators only reaches the CRLB for much higher SNRs and the variance steadily increases as the SNR is lowered; however, the low-pass filters do not need to deal with the problem of angle unwrapping. The latter class of estimator, with the low-pass filters in Table 2, was used to re-process signal type 2 with configuration 1 (see Section 6.2.1). These estimators are arguably more robust because failure (at low SNR) is replaced by a more gradual performance degradation (see Figure 40). In these estimators, the low-pass filters must track a complex sinusoid instead of a real polynomial thus wider bandwidth filters are generally required. The low-pass filters described in this paper, designed using derivative constraints at dc for unbiased polynomial tracking at steady state, perform reasonably well; however, other design techniques may offer better performance for this alternative class of estimator.

The FIR and IIR filters designed here are well suited to the problem of instantaneous phase and frequency estimation; however, their low-pass responses make them suitable for a wide range of other applications that require (fixed-lag) smoothers and predictors, to track a low-frequency (polynomial-like) signal in high-frequency noise. The $K_\varphi$th-order IIR filters are particularly attractive in applications that require a long timescale (thus a narrow bandwidth) because they have a lower complexity than their FIR equivalents, and the proposed design procedure is simpler than many other methods. The bandwidth of the IIR filters is configurable (via the $f_c$ parameter), the high frequency attenuation is adjustable (via the $K_0$ parameter), and the passband flatness (phase and magnitude) is controllable (via the $K_1$ parameter). IIR filters are often dismissed due to their lack of perfect phase linearity; however, the deviation from phase linearity over the passband is less than 3 degrees for all IIR estimators considered here, as shown in Figure 10 & Figure 26 (and also for the filters in Table 3 & Table 4). Note that both FIR and IIR predictors have non-linear phase responses, by design.

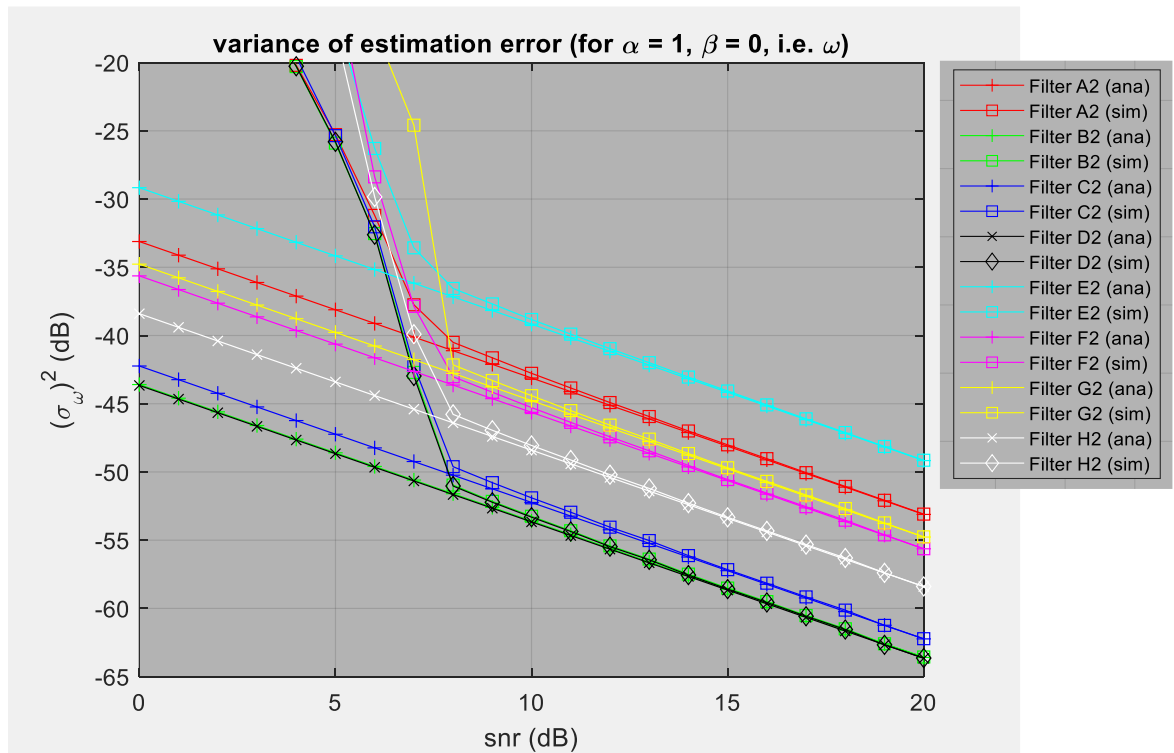

Figure 34. Error variance versus SNR for signal type 2 and system configuration 1 using the FIR & IIR filters in Table 2, with uniformly distributed measurement noise (instead of Gaussian noise).

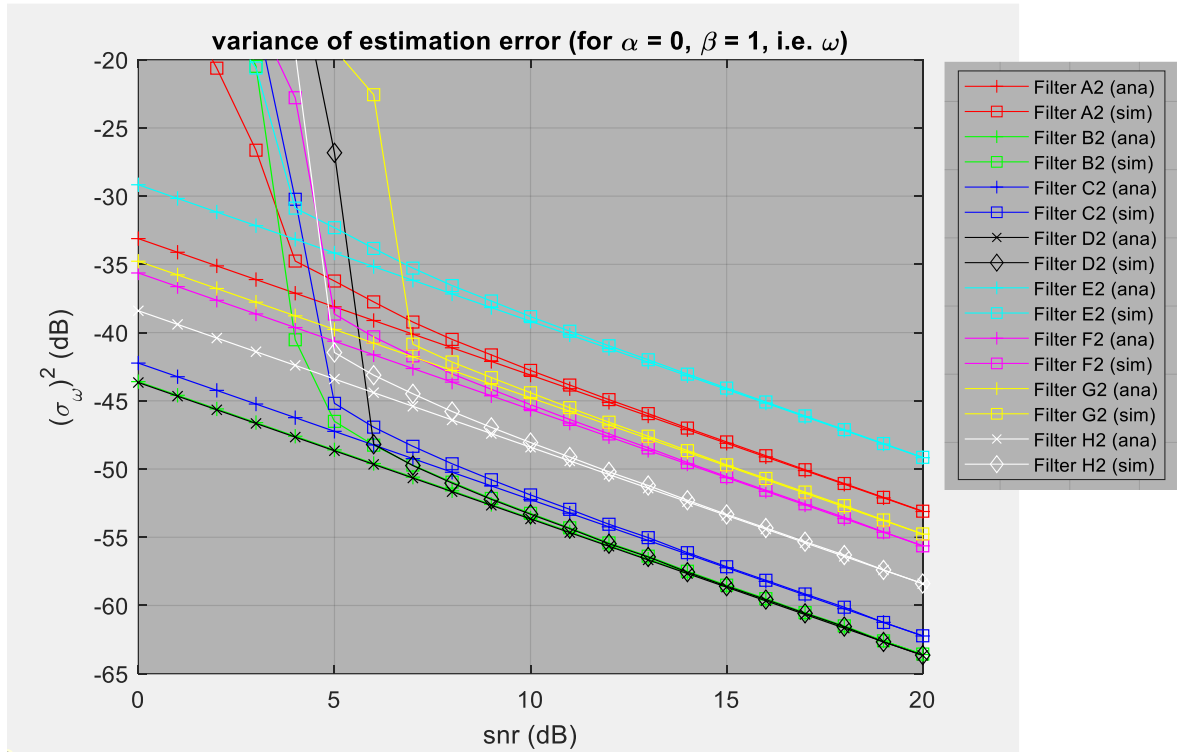

Figure 35. Error variance versus SNR for signal type 2 and system configuration 2 using the FIR & IIR filters in Table 2, with uniformly distributed measurement noise (instead of Gaussian noise).

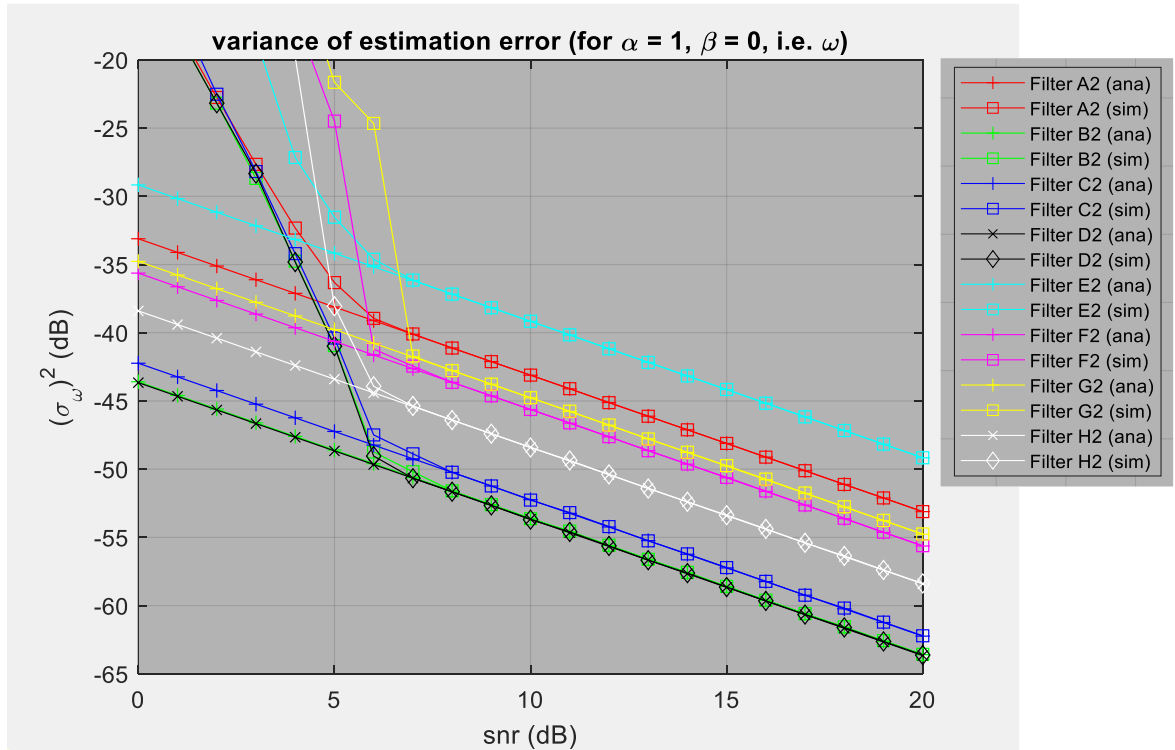

Figure 36. Error variance versus SNR for signal type 2 and system configuration 1 using the FIR & IIR filters in Table 2, with angle measurement noise added (instead of complex noise).

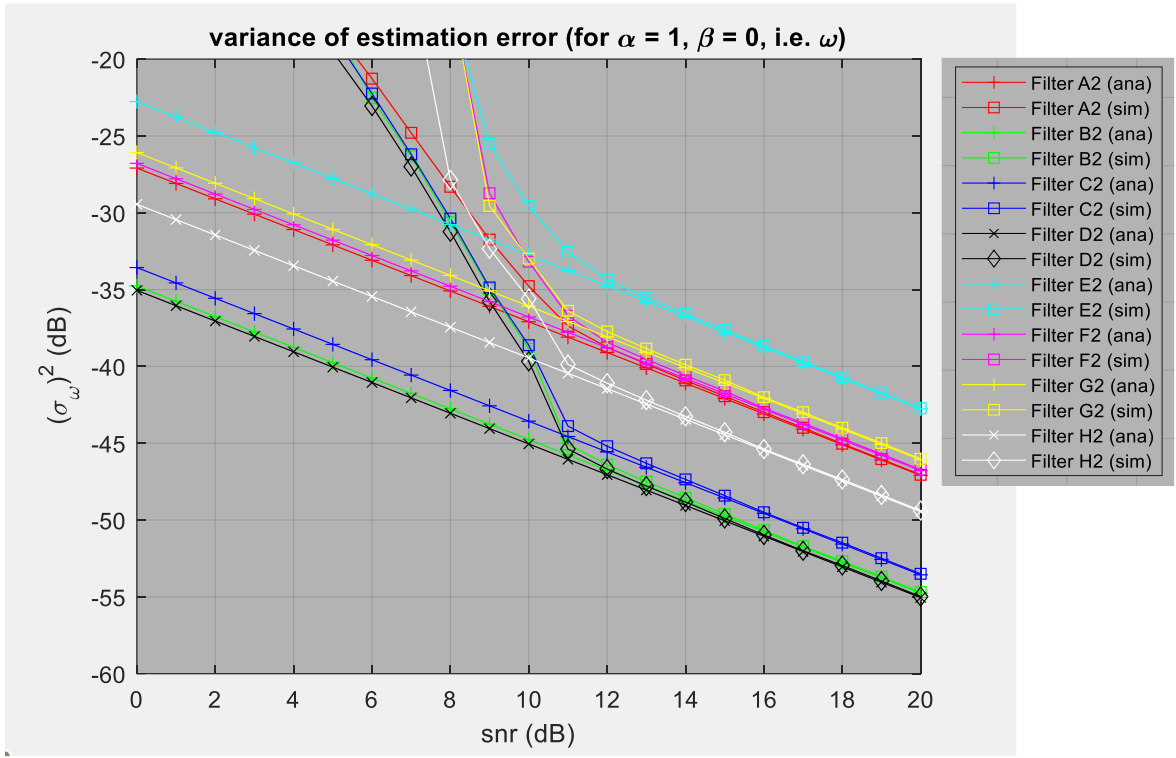

*Figure 37. Error variance versus SNR for signal type 2 and system configuration 1 using the FIR & IIR filters in Table 2 with M = 32 instead of M = 64.*

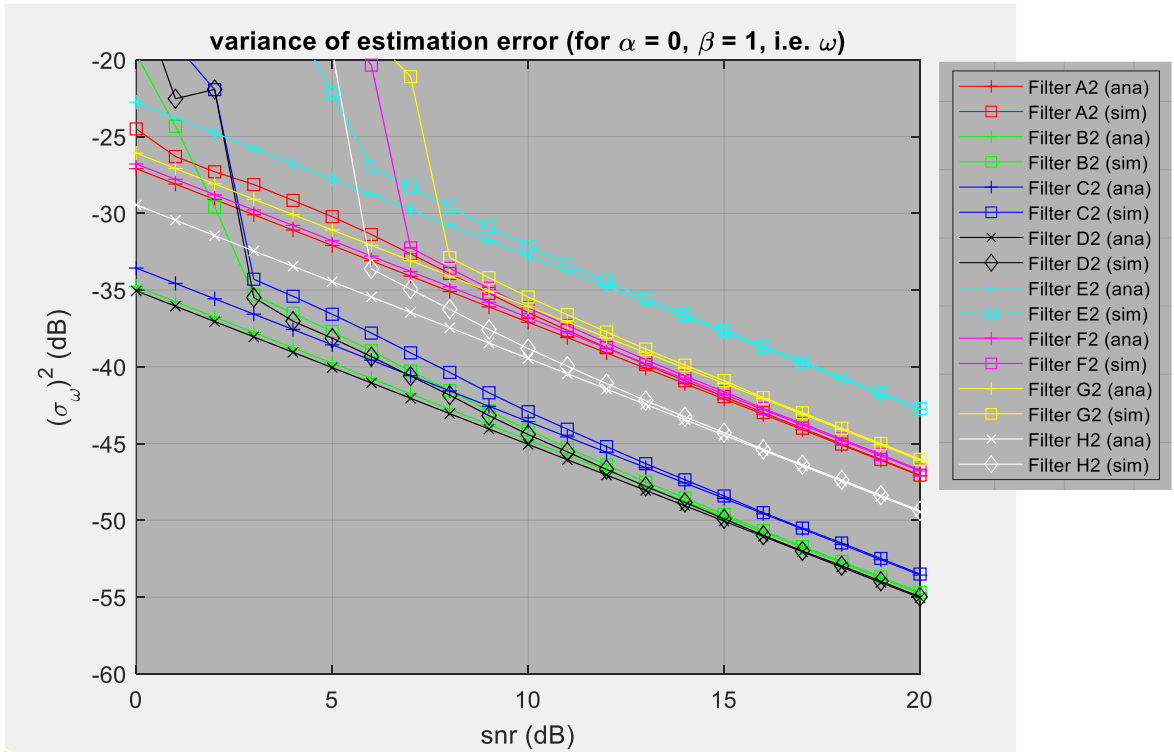

*Figure 38. Error variance versus SNR for signal type 2 and system configuration 2 using the FIR & IIR filters in Table 2 with M = 32 instead of M = 64.*

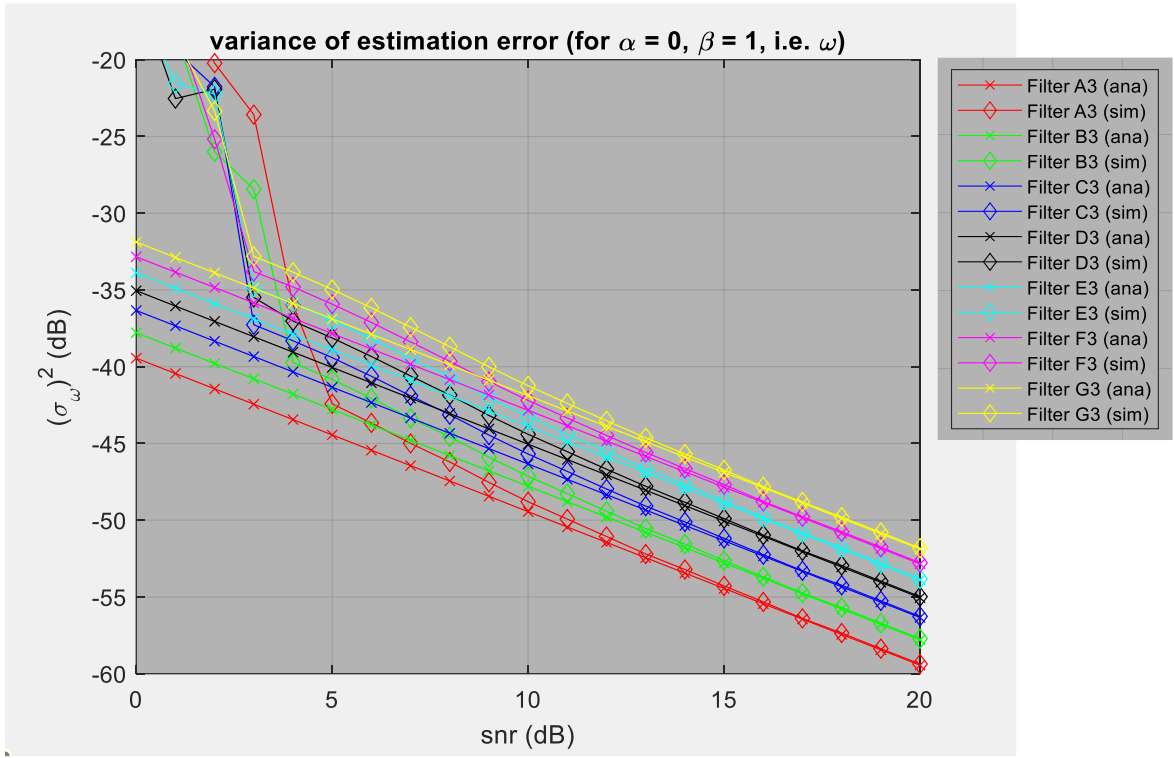

*Figure 39. Error variance versus SNR for signal type 2 and system configuration 2 using the IIR filters in Table 3 with $M = 32$ instead of $M = 64$.*

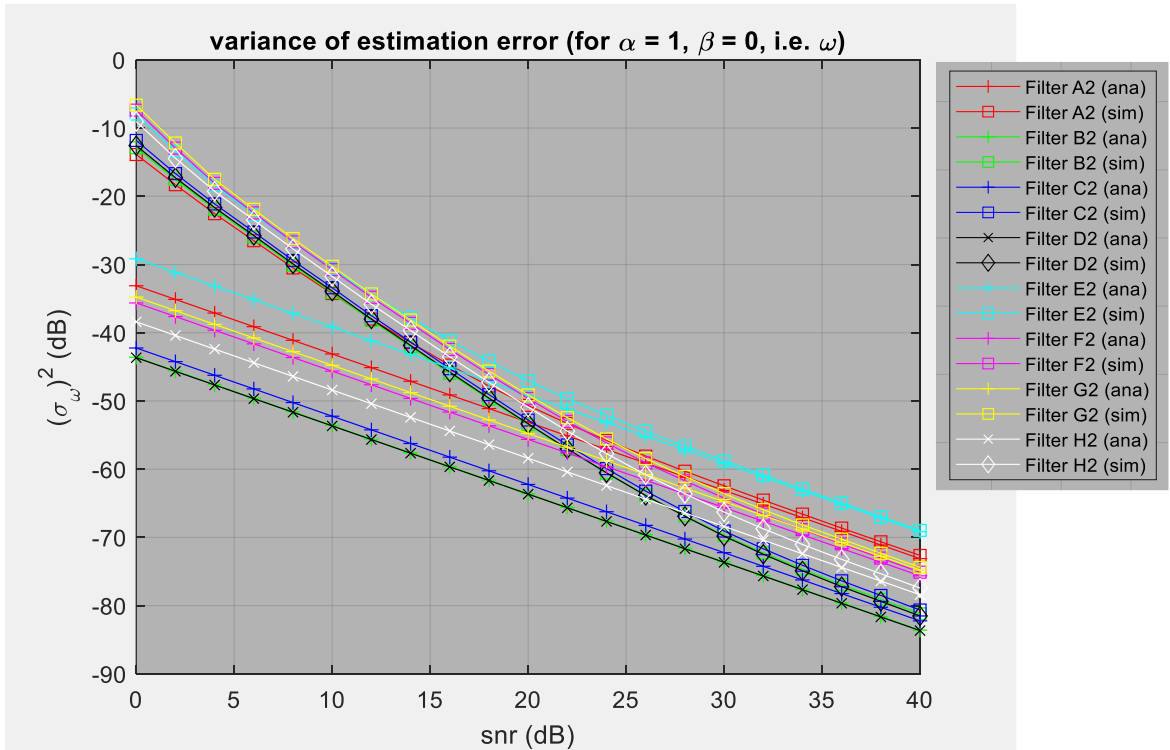

*Figure 40. Error variance versus SNR for signal type 2 and system configuration 1 using the FIR & IIR filters in Table 2, applied in the complex domain (instead of the angle domain).*

# 8 Conclusion

Re-casting the instantaneous frequency estimation problem as an instantaneous phase estimation problem provides several potential benefits in sensor signal-processing systems. Firstly, estimates of phase offsets become available, e.g. for measurements of time delay and radial distance. Secondly, precise and accurate estimates of frequency, e.g. for measurements of Doppler shift and radial velocity, that reach the theoretical lower bound, are available down to a significantly lower SNR. Well-designed sequential smoothers and predictors, with non-recursive or recursive structures, are shown here to be essential for the realization of these benefits in online processors with streaming architectures. The (non-recursive) FIR case is well studied in the literature; therefore, this paper deals mainly with the (recursive) IIR case. The proposed filter design procedure utilizes the relationship between low-order derivatives at the origin of the frequency domain (i.e. at $\omega = 0$) and low-order moments around the local phase origin in the time domain (i.e. at $n - q$). The resulting filters therefore have a low-pass frequency response with a configurable bandwidth (thus timescale). Responses with a narrower bandwidth are more precise at high SNR (greater noise attenuation), whereas responses with a wider bandwidth are more accurate at low SNR (fewer unwrapping errors). The FIR and IIR estimators may be configured for very similar performance; however, narrowband IIR structures have a lower computational complexity than their FIR equivalents. The simulations demonstrate (in principle) that narrowband filters may be used to process swept-frequency signals with a bandwidth that is many times greater than the filter and the sampler, provided the estimator has sufficient time to establish an angle track, before the rate of phase change becomes large.

The use of a prediction filter with a one-sample phase lead is shown (using MC simulation) to greatly reduce the incidence of angle unwrapping errors when estimating the instantaneous frequency of a complex exponential using a smoothing filter. For this problem, a (high-pass) differentiator is usually placed *before* the (low-pass) smoothing filter with a multi-sample phase lag to minimize the error variance of the frequency estimate. The differentiator reduces the rate of angle change that the smoother must handle, which reduces the incidence of angle unwrapping errors, that limit performance at low SNR, resulting in a failure "threshold" at around 10 dB. The prediction filter allows a higher rate of angle change to be tracked without unwrapping errors, thus the differentiator may be placed *after* the smoothing filter. As the differentiator amplifies high-frequency noise (linearly at high SNR and non-linearly at low SNR) phase unwrapping is accomplished in a more benign noise environment; however, the bandwidth of the filter must be sufficient to ensure that it converges before angle wrapping events occur. The filter bandwidth is clearly a critical parameter and tuning guidelines are provided for both FIR and IIR filters. Using the predictor and estimator (i.e. the smoother) in tandem lowers the SNR threshold below 10 dB, down to 3dB in some cases. Moreover, removing the pre-differentiation stage allows the instantaneous phase to be estimated, in addition to the instantaneous frequency, using either (non-recursive) FIR or (recursive) IIR estimators.

A novel procedure for the design of recursive IIR filters, with an optimal passband group-delay, and good passband phase linearity, is described. The IIR filters have a lower computational complexity than their FIR equivalents. The proposed procedure is relatively simple because the positions of the filter poles are pre-determined, for the desired filter bandwidth. Furthermore, it does not rely upon prior knowledge of process-noise, measurement-noise, or initial-state, probability density functions. The FIR and IIR filters are designed and used here to estimate the instantaneous phase and frequency of exponential signals with polynomial phase; however, they may also be useful in other applications that require low-pass estimators (i.e. smoothers) and predictors for polynomial-like signals in discrete time.